\documentclass[12pt,leqno]{amsart}
\usepackage{amsmath,epsfig,graphicx,color, float}
\usepackage{subcaption}
\usepackage{relsize}
\usepackage{comment}
\usepackage{hyperref}
\usepackage{pdfpages}
\usepackage{graphicx}
\usepackage{mwe}
\usepackage{algorithm}
\usepackage{algorithmicx}
\usepackage{algpseudocode}
\usepackage{natbib}

\usepackage[margin=1.2in]{geometry}

\numberwithin{table}{section}
\numberwithin{figure}{section}
\numberwithin{equation}{section}

\definecolor{darkblue}{rgb}{.2, 0.2,.8}
\definecolor{darkgreen}{rgb}{0,0.5,0.3}
\definecolor{darkred}{rgb}{.8, .1,.1}

\newcommand{\bfX}{\vect{X}}

\newcommand{\bfT}{\mat{T}}
\newcommand{\bft}{\vect{t}}
\newcommand{\bfe}{\vect{e}}
\newcommand{\bfpi}{\vect{\pi}^\mathsf{T}}
\newcommand{\bfp}{\vect{\pi}}

\newcommand{\0}{\mat{0}}

\renewcommand{\P }{{\mathbb P}}

\newcommand{\ex}{{\rm e}}

\newtheorem{lemma}{Lemma}[section]

\newtheorem{proposition}[lemma]{Proposition}

\newtheorem{example}[lemma]{Example}

\newtheorem{remark}{Remark}[section]

\usepackage{bm}
\newcommand{\vect}[1]{\pmb{#1}}
\newcommand{\mat}[1]{\boldsymbol{\bm #1}}

\DeclareMathOperator*{\argmax}{arg\,max}

\usepackage{marginnote}

\allowdisplaybreaks

\begin{document}

\bibliographystyle{apalike}
\title{Phase-type distributions for claim severity regression modeling}

\author[M. Bladt]{Martin Bladt}
\address{Department of Actuarial Science, Faculty of Business and Economics, University of Lausanne, UNIL-Dorigny,
1015 Lausanne}
\email{martin.bladt@unil.ch}

\begin{abstract}
{   This paper addresses the task of modeling severity losses using segmentation when the data distribution does not fall into the usual regression frameworks. This situation is not uncommon in lines of business such as third-party liability insurance, where heavy-tails and multimodality often hamper a direct statistical analysis. We propose to use regression models based on phase-type distributions, regressing on their underlying inhomogeneous Markov intensity and using an extension of the EM algorithm. These models are interpretable and tractable in terms of multi-state processes and generalize the proportional hazards specification when the dimension of the state space is larger than one. We show that the combination of matrix parameters, inhomogeneity transforms, and covariate information provides flexible regression models that effectively capture the entire distribution of loss severities.}

\end{abstract}
\subjclass{Primary; Secondary }
\maketitle

\section{Introduction}

The task of modeling claim severities is a well-known and difficult challenge in actuarial science, and their correct description is of large interest for the practicing actuary or risk manager. In some lines of business, the size of claims may manifest features which are difficult to capture correctly by simple distributional assumptions, and ad-hoc solutions are very often used to overcome this practical drawback. In particular, heavy-tailed and multi-modal data may be present, which together with a high count of small claim sizes, produces an odd-shaped distribution.

In this paper we introduce phase--type (PH) distributions as a powerful vehicle to produce probabilistic models for the effective description of claim severities. PH distributions are roughly defined as the time it takes for a pure-jump Markov processes on a finite state space to reach an absorbing state. Their interpretation in terms of traversing several states before finalizing is particularly appealing actuarial science, since one may think of claim sizes as the finalized ``time" after having traversed some unobserved states to reach such magnitude, for instance legal cases, disabilities, unexpected reparation costs, etc. Many distributions such as the exponential, Erlang, Coxian and any finite mixture between them fall into PH domain, and they possess a universality property in that they are dense (in the sense of weak convergence, cf. \cite{asmussen2008applied}) on the set of probability measures with positive support.
 
Despite many particular instances of PH being classical distributions, a unified and systematic approach was only first studied in \cite{neuts75,neuts1981matrix},  which laid out the modern foundations of matrix analytic methods (see \cite{Bladt2017} for a recent comprehensive treatment). Since then, they have gained popularity in applied probability and statistics, due to their versatility and closed-form formulas in terms of matrix exponentials. A key development for their use in applied statistics was the development of a maximum likelihood estimation procedure via the expectation-maximization (EM) algorithm (cf.  \cite{asmussen1996fitting}), which considers the full likelihood arising from the path representation of a phase--type distribution. { \cite{ahn2012new} explore the case of exponentially-transformed PH distributions to generate heavy-tailed distributions.} Furthermore, transforming PH distributions parametrically has been considered in  \cite{albrecher2019inhomogeneous, albrecher2020iphfit}, resulting in inhomogeneous phase-type distributions (IPH), which leads to non-exponential tail behaviors by allowing for the underlying process to be time-inhomogeneous (see also \cite{bladt2017fitting,mml} for alternative heavy-tailed PH specifications). The latter development allows to model heavy- or light-tailed data with some straightforward adaptations to the usual PH statistical methods.

Incorporating rating factors, or covariates, is particularly relevant when segmentation of claims is sought for, e.g. for insurance pricing, and is known to be problematic when the underlying model does not belong to the exponential dispersion family (GLM). Regression models for PH distributions have been considered in the context of survival analysis, cf. \cite{mcgrory2009fully} and references therein for the Coxian PH case, and more recently \cite{bladt2021semi} for IPH distributions {  with applications to mortality modeling}. {  The current work uses the specification of the model within the latter contribution but applied to claim severity modeling for insurance. The methodology is simpler, since we treat the fully observed case, allowing for some closed-form formulas for e.g. mean estimation or inferential tools.}

Several other statistical models for insurance data can be found in the literature, with the common theme being constructing a global distribution consisting of smaller simpler components, either by mixing, splicing (also referred to as composite models), or both. In \cite{lee2010modeling}, a mixture of Erlang distributions with common scale parameter was proposed, and subsequently extended to more general mixture components in \cite{tzougas2014optimal, miljkovic2016modeling, fung2020new}. Concerning composite models, many different tail and body distribution combinations have been considered, and we refer to \cite{grun2019extending} for a comparison of some of them, where a substantial body of the relevant literature can be found as well. Global models is dealt with a combination of mixtures and splicing, with \cite{reynkens2017modelling} being the main reference (see also \cite{fung2021mixture} for a feature-selection approach). 

Our present PH regression model is not only mathematically but also conceptually different to all previous approaches. With the use of matrix calculus, it allows to specify multi-modal and heavy-tailed distributions without the need of threshold selection or having to specify the number of mixture components. Instead, an inhomogeneity function and the size of the underlying state space have to be specified. The former is usually straightforward according to the tail behaviour of the data, and the latter can be chosen either in a data-driven way or in a more dogmatic way if there is a certain belief of a specific number of unobserved states driving the data-generating process.

The structure of the remainder of the paper is as follows. In Section \ref{sec:pre}, we provide the necessary preliminaries on IPH distributions { and formulate our PH regression model including their parametrizations, their estimation procedure, statistical inference, and dimension and structure selection. We then perform a simulation study in Section \ref{sec:sim} to illustrate its effectiveness in a synthetic heterogeneous dataset. Subsequently, we perform the estimation on a real-life French insurance dataset in Section \ref{sec:reg}. Finally, Section  \ref{sec:conclusion} concludes.
}

\section{Model specifications}\label{sec:pre}
In this section we {  lay down} some preliminaries on inhomogeneous phase--type distributions which are needed { in order to introduce} the relevant the key concepts to be used in our regression, {  and subsequently proceed to introduce the specifications of the claim severity models}. We skip the proofs, and the interested reader can find a comprehensive treatment { of the marginal models} in the following references: \cite{Bladt2017,albrecher2019inhomogeneous}, the latter taking a more abstract approach. We will however, tailor and emphasize the most relevant features of the models with respect to { claim severity modeling for }insurance.

\subsection{Mathematical formulation of inhomogeneous phase--type distributions}
Let $ ( J_t )_{t \geq 0}$ be a time--inhomogeneous Mar\-kov pure-jump process on the finite state space $\{1, \dots, p, p+1\}$, where states $1,\dots,p$ are transient and $p+1$ is absorbing. In our setting, such a process can be regarded as an insurance claim evolving through different states during its lifetime until the absorbing state is reached, which determines the total size of the claim. {  In its most general form, the transition probabilities of the jump process
\begin{align*}
p_{ij}(s,t)=\P(J_t=j|J_s=i),\quad i,j\in\{1,\dots,p+1\},
\end{align*}
may be written in succinct matrix form in terms of the product integral as follows:

$$\mat{P}(s,t)=\prod_{s}^{t}(\boldsymbol{I}+\boldsymbol{\Lambda}(u) d u):=\boldsymbol{I}+\sum_{k=1}^{\infty} \int_{s}^{t} \int_{s}^{u_{k}} \cdots \int_{s}^{u_{2}} \mathbf{\Lambda}\left(u_{1}\right) \cdots \mathbf{\Lambda}\left(u_{k}\right) d u_{1} \cdots \mathrm{d} u_{k},$$
for $s<t,$ where $\mat{\Lambda}(t)$ is a matrix with negative diagonal elements, non-negative off-diagonal elements and rows sums equal to zero, commonly referred to as an intensity matrix. Since the process $ ( J_t )_{t \geq 0}$ has a decomposition in terms of transient and absorbing states, we may write}
\begin{align*}
	\mat{\Lambda}(t)= \left( \begin{array}{cc}
		\bfT(t) &  \bft(t) \\
		\0 & 0
	\end{array} \right)\in\mathbb{R}^{(p+1)\times(p+1)}\,, \quad t\geq0\,,
\end{align*}
where $\bfT(t) $ is a $p \times p$ sub-intensity matrix and $\bft(t)$ is a $p$--dimensional column vector providing the exit rates from each state directly to $p+1$. The intensity matrix has this form since there are no positive rates from the absorbing state to the transient ones, and  since the rows should sum to zero, we have that for $t\geq0$, $\bft (t)=- \bfT(t) \, \bfe$, where $\bfe $ is a $p$--dimensional column vector of ones. Thus, $\bfT(t)$ is sufficient to describe the dynamics of process after time zero. We write $\bfe_k$ in the sequel for the $k$-th canonical basis vector of $\mathbb{R}^p$, so in particular $\bfe=\sum_{k=1}^p\bfe_k$.

If the matrices $\bfT(s)$ and $\bfT(t)$ commute for any $s<t$ we may write 
\begin{align*}
\mat{P}(s,t)
=\begin{pmatrix}
    \exp\left(\int_s^t\bfT(u)du\right) & \bfe-\exp\left(\int_s^t\bfT(u)du\right) \bfe \\
    \textbf{0} & 1
    \end{pmatrix},\quad s<t.
\end{align*}

The matrices $(\bfT(t))_{t\ge0}$ together with the initial probabilities of the Markov process, $ \pi_{k} = \P(J_0 = k)$, $k = 1,\dots, p$, or in vector notation $$\bfp = (\pi_1 ,\dots,\pi_p )^\mathsf{T},$$  fully parametrize inhomogeneous phase-type distributions, and with the additional assumption that $\P(J_0 = p + 1) = 0$ we can guarantee that absorption happens almost surely at a positive time. We thus define the positive and finite random variable given as the absorption time as follows
\begin{align*}
	Y = \inf \{ t >  0 : J_t = p+1 \},
\end{align*}
and say that it follows an inhomogeneous phase--type distribution with representation $(\bfp,(\bfT(t))_{t\ge0})$. For practical applications, however, we focus on a narrower class which allows for effective statistical analysis. We make the following assumption regarding the inhomogeneity of the sub-intensity matrix:
$$\bfT(t) = \lambda(t)\,\bfT,$$ with $\lambda(t)$ some parametric and positive function such that the map $$y\mapsto\int_0^y \lambda(s)ds\in(0,\infty),\quad\forall y>0,$$ converges to infinity as $y\to\infty$, and $\bfT$ is a sub--intensity matrix that does not depend on time $t$. When dealing with particular entries of the sub-intensity matrix, we introduce the notation
$$\bfT=(t_{kl})_{k,l=1,\dots,p},\quad \bft=-\bfT\bfe=(t_1,\dots,t_p)^{\mathsf{T}},\quad \alpha_i=-t_{ii}>0,\:\:\: i=1,\dots,p.$$

We then write $$Y \sim  \mbox{IPH}(\bfp , \bfT , \lambda ),$$ where IPH stands for inhomogeneous phase-type. Similarly, we write PH in place of phase-type in the sequel, i.e. for the case when $\lambda$ is a constant.

IPH distributions are the building blocks to model claim severity for two reasons. Firstly, they have the interpretation of being absorption times of an underlying unobserved multi-state process evolving through time, which can be very natural if the losses are linked with an evolving process, such as a lifetime or a legal case; and secondly, PH distributions are the archetype generalization of the exponential distribution in that many closed-form formulas are often available in terms of matrix functions, and thus, when armed with matrix calculus and a good linear algebra software, a suite of statistical tools can be developed explicitly. 

It is worth mentioning that PH distributions possess a universality property since they are known to be dense on the set of positive-valued random variables in the sense of weak convergence. Consequently, they possess great flexibility for atypical histogram shapes that the usual probabilistic models struggle to capture. Do note however, that the tail can be grossly misspecified if no inhomogeneity function is applied, since weak convergence does not guarantee tail equivalence between the approximating sequence and the limit.

Another feature of IPH distributions is that they can have a different tail behaviour than their homogeneous counterparts, allowing for non-exponential tails. To see this, first note that if $Y \sim  \mbox{IPH}(\bfp , \bfT , \lambda )$, then we may write
 \begin{equation}\label{gtrans}
Y \sim g(Z) \,,
\end{equation}where $Z \sim \mbox{PH}(\bfp , \bfT )$ and $g$ is defined in terms of $\lambda$ by
\begin{equation*}
g^{-1}(y) = \int_0^y \lambda (s)ds,\quad y\ge0,   \label{eq:transformation-g}
\end{equation*}
which is well-defined by the positivity of $\lambda$, or equivalently (see for instance Theorem 2.9 in \cite{albrecher2019inhomogeneous})
\begin{equation*}
\lambda (s) = \frac{d}{ds}g^{-1}(s) \,.  \label{eq:transformation-lambda}
\end{equation*}

Since the asymptotic behaviour of PH distributions can be deduced from its Jordan decomposition, the following result in the inhomogeneous case is available. It is a generalization of the theory developed in \cite{bladt2021semi}, and the proof is hence omitted.
\begin{proposition}\label{prop:functionals}
	Let $ Y \sim\mbox{IPH}(\bfp , \bfT , \lambda )$. Then the survival function { $S_Y=1-F_Y(y)$}, density $f_Y$, hazard function $h_Y$ and cumulative hazard function $H_Y$ of $Y$ satisfy, respectively, as $t\to \infty$,
	\begin{align*}
	S_Y(y)&= {\bfpi}\exp \left( \int_0^y \lambda (s)ds\ \mat{T} \right)\vect{e} \sim c_1 [g^{-1}(y)]^{n -1} e^{-\chi [g^{-1}(y)]},\\
f_Y(y) &=\lambda (y)\, {\bfpi}\exp \left( \int_0^y \lambda (s)ds\ \mat{T} \right)\vect{t}\sim c_2 [g^{-1}(y)]^{n -1} e^{-\chi [g^{-1}(y)]}\lambda(y),\\
		h_Y(y) &\sim c \lambda(y) ,\\
		H_Y(y)& \sim k g^{-1}(y),
	\end{align*}
	where $c_1,c_2,c,k$ are positive constants, $-\chi$ is the largest real eigenvalue of $\bfT$ and $n$ is the dimension of the Jordan block associated to $\chi$.
\end{proposition}

In particular we observe that the tail behaviour is determined almost entirely by function $g$. We also note that any model based on the intensity of the process (as will be the case for our regression model below) is asymptotically a model on the hazard of the distribution.

If $g$ is the identity, we get Erlang tails, but for other choices, we move away from the exponential domain. { The choice of $g$ is typically determined according to the desired tail behaviour required for applications. For heavy-tailed claim severities in the Fr\'echet max-domain of attraction, the exponential transformation $g(z)=\eta \left( \exp(z)-1\right),\:\:\eta>0$ leading to Pareto tails is the simplest choice, although the log-logistic can also have such a behaviour. When data is moderately heavy-tailed, as understood by being in the Gumbel max-domain of attraction, heavy-tailed Weibull $g(z)=z^{1/\eta},\:\:\eta>0$, or Lognormal $g(z)=\exp(z^{1/\gamma})-1,\:\:\gamma>1$ transforms can provide useful representations. In life insurance applications the Matrix-Gompertz distribution with $g(z)=\log( \eta z  + 1 ) / \eta,\:\: \eta>0$ can be particularly useful to model human lifetimes.}


One tool which is particularly useful to deal with expressions involving IPH distributions is matrix calculus. Here, the main tool is Cauchy's formula for matrices, which is as follows. If $u$ is an analytic function and $\mat{A}$ is a matrix, we define
\begin{align*}
	u( \mat{A})=\dfrac{1}{2 \pi i} \oint_{\Gamma}u(w) (w \mat{I} -\mat{A} )^{-1}dw ,\quad \mat{A}\in \mathbb{R}^{p\times p}.
\end{align*}
where $\Gamma$ is a simple path enclosing the eigenvalues of $\mat{A}$, and $\mat{I}$ is the $p$-dimensional identity matrix. This allows for functions of matrices to be well defined, which in turn gives explicit formulas for IPH distributions, simplifying calculations and numerical implementation alike. 

%

\begin{example}\normalfont \label{densities_example1}
Consider the following underlying Markov process parameters
$$\bfp=(1,\,0,\,0)^{\mathsf{T}},\quad
\bfT=\begin{pmatrix}
-100 & 50 & 0 \\
0 & -1 & 1/2\\
0 & 0 & -1/100 \\
\end{pmatrix},$$
and a Weibull inhomogeneity parameter $\eta=8$, which was chosen significantly light-tailed for visualisation purposes. Compared to a conventional Weibull distribution, with $\bfp=1$, $\bfT=-1/10$ and $\eta=8$, we see in Figure \ref{ex:weibull_dens} that just a few additional parameters in the augmented matrix version of a common distribution can result in very different behaviour in the body of the distribution (the tail behaviour is guaranteed to be the same). For this structure, the case $p=3$ shows that the Markov process can traverse $3$ states before absorption, resulting in three different modes of the density. For other matrix structures or parameters, there is in general no one-to-one correspondence between $p$ and the number of modes of the resulting density.

\begin{figure}[!htbp]
\centering
\includegraphics[width=0.7\textwidth]{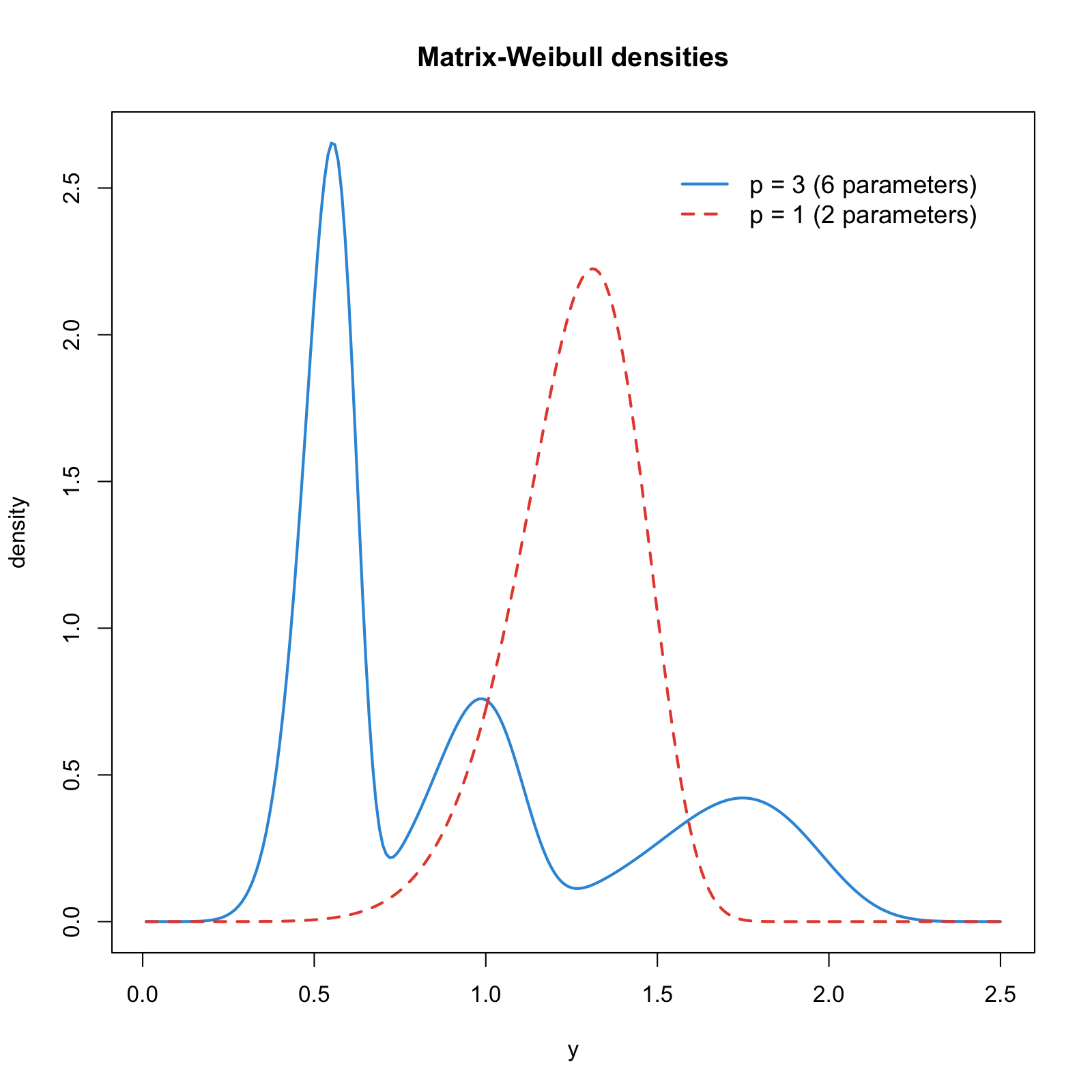}
\caption{Densities corresponding to a scalar Weibull distribution and a Matrix-Weibull distribution.
} \label{ex:weibull_dens}
\end{figure}

\end{example}

\subsection{The Phase-type regression model}\label{sec:prop_int}
We now introduce a regression model based on IPH distributions and use it to provide a segmentation model for claim severities. Classically, for pure premium calculation, the mean is the main focus. {  Presently, we do not aim to improve on other data-driven procedures for pure mean estimation but instead focus on the entire distribution of the loss severities. That said, a consequence of a well-fitting global model is a reasonable mean estimate in the small sample case. However, when considering large sample behavior, any scoring rule in terms of means is not a proper scoring rule for target distributions that are not fully characterized by their first moment (see \cite{gneiting2007strictly}), and thus model choice is less important in this case.}

Let $\bfX = (X_1, \dots, X_d)^\mathsf{T}$ be a $d$ -dimensional vector of rating factors associated with the loss severity random variable $Y$. We keep the convention that $\bfX$ does \textit{not} contain the entry of $1$ associated with an intercept, in order to obtain an identifiable regression specification (the intercept is included in the matrix $\bfT$). Let $\vect{\beta}$ be a $d$-dimensional vector of regression coefficients, so that $\bfX^\mathsf{T}\vect{\beta}$ is a scalar. Then we define the conditional distribution of $Y|\mat{X}$ as having density function, see Proposition \ref{prop:functionals},
 \begin{eqnarray}\label{iph_density}
 f(y) &=& m(\bfX^\mathsf{T}\vect{\beta}) \lambda (y;  \theta )\, {\bfpi}\exp \left( m(\bfX^\mathsf{T}\vect{\beta})\int_0^y \lambda (s;  \theta )ds\ \mat{T} \right)\vect{t} \,, \label{eq:dens-prop_int}
\end{eqnarray}
where $m:\mathbb{R}\to \mathbb{R}_+$
is any positive function (monotonicity is not required). We call this the \textit{phase-type regression model}. In particular we recognise an IPH distribution $ \mbox{IPH}(\bfp , \bfT , \lambda(\cdot \mid  \bfX , \vect{\beta}, \theta  ) )$, where

\begin{align}\label{prop_intens_spec}
	\lambda(t \mid  \bfX ,  \vect{\beta},\theta  ):=\lambda ( t ;   \theta  ) m(\bfX^\mathsf{T}\vect{\beta})  \,.
\end{align}

One possible interpretation of this model is that covariates act multiplicatively on the underlying Markov intensity, thus creating proportional intensities among different policies. In fact, since the intensity of an IPH distribution is asymptotically equivalent to its hazard function, we have that covariates satisfy a generalized proportional hazards specification in the far-right tail.

The conditional mean can be written in the form
\begin{align}\label{mean_equation}
\mu(Y|\bfX)=\int_0^\infty \bfpi\exp\left(m(\bfX^\mathsf{T}\vect{\beta})\int_0^y \lambda(s; \theta )ds\,\bfT \right)\bfe \,dy,
\end{align}
which is simply obtained by integrating the survival function over $\mathbb{R}_+$, see Proposition \ref{prop:functionals}. A simple special case is obtained by the following choices, giving a Gamma GLM with canonical link: take $\bfT=-1$, and $\lambda\equiv 1$ to receive
\begin{align*}
\mu(Y|\bfX)=\int_0^\infty \exp(-m(\bfX^\mathsf{T}\vect{\beta}) y ) \,dy=\frac{1}{ m(\bfX^\mathsf{T}\vect{\beta})}.
\end{align*}

Another slightly more complex special case is that of regression for Matrix-Weibull distributions, which contains the pure PH specification (when $\lambda\equiv1$). In this setting it is not hard to see that
\begin{align}\label{weibull_mean_expression}
\mu(Y|\bfX)=\int_0^\infty \bfpi\exp\left(m(\bfX^\mathsf{T}\vect{\beta})\bfT y^{ \theta }\right)\bfe \,dy =\frac{\Gamma(1+ \theta ^{-1}) \bfpi\bfT^{- \theta ^{-1}}\bfe}{ m(\bfX^\mathsf{T}\vect{\beta})^{ \theta ^{-1}}}.
\end{align}
 
\begin{figure}[!htbp]
\centering
\includegraphics[width=0.7\textwidth]{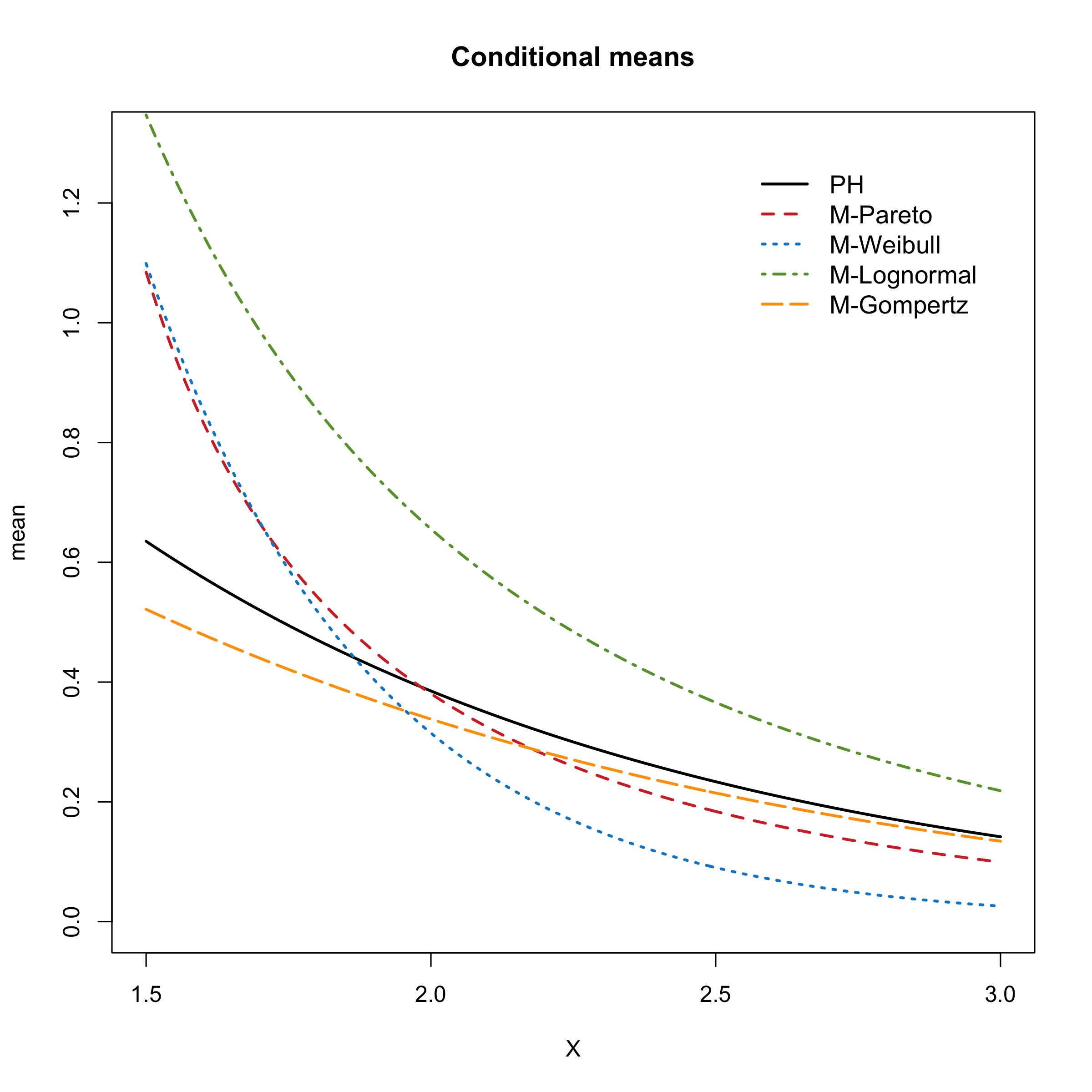}
\caption{Mean functions of the PH regression model, as a function of a univariate regressor and for different specifications of the inhomogeneity function $\lambda$.
} \label{means}
\end{figure}

Figure \ref{means} shows \eqref{mean_equation} as a function of a unidimensional $\bfX$ for various inhomogeneity functions $\lambda$, and fixing the underlying Markov parameters at $\bfp=(1/2,1/4,1/4)^{\mathsf{T}}$ and
$$
\bfT=\begin{pmatrix}
-1 & 1/2 & 1/4 \\
1/4 & -1 & 1/4\\
1/4 & 1/4 & -1 \\
\end{pmatrix}.$$

\subsection{Estimation}
{  Claim severity fitting can be done by maximum likelihood estimation (MLE) in much the same way whether there are rating factors or not, and hence we exclusively present the more general case. For this purpose, we adopt a generalization of the expectation-maximization (EM) algorithm, which we outline below. Such an approach is much faster to converge than naive multivariate optimization, provided a fast matrix exponential evaluation is available (see \cite{moler1978nineteen}).}

A full account of the EM algorithm for ordinary PH distributions can be found in \cite{asmussen1996fitting} which led to the subsequent developments and extensions to censored and parameter dependent transformations in \cite{olsson1996estimation, albrecher2020iphfit}. Further details are provided in Appendix \ref{apA}, since the formulas therein are the building blocks of { the EM algorithm for the PH regression model}.

By defining $g(\, \cdot \,; \theta )$ and $g(\, \cdot \,|\, \bfX, \vect{\beta}, \theta )$ as follows in terms of their inverse functions $g^{-1}(y;  \theta )=\int_{0}^{y}\lambda(s;  \theta )ds$ and $$g^{-1}(y \,|\, \bfX, \vect{\beta},  \theta )=\int_{0}^{y}\lambda(s \,|\, \bfX,  \theta )ds = g^{-1}(y;  \theta ) \exp(\bfX^\mathsf{T}\vect{\beta} ),$$ we note the simple identity
\begin{align}\label{PHRrep}
	g(y \,|\, \bfX,   \vect{\beta},\theta  )=g(y \exp( -\bfX^\mathsf{T}\vect{\beta} ) \,;\,  \theta  )   \,,
\end{align}
which yields that $g^{-1}( Y \,|\, \bfX,  \vect{\beta}, \theta ) \sim \mbox{PH}(\bfp , \bfT )$. In other words, there is a parametric transformation, depending both on the covariates and regression coefficients, which brings the PH regression model into a conventional PH distribution. The generalized EM algorithm hinges on this fact, and optimizes the transformed PH distribution and the transformation itself in alternating steps.

\begin{algorithm}[]
\caption{Generalized EM algorithm for phase-type regression}\label{alg;IPHreg}
\begin{algorithmic}
\State \textit{\textbf{Input}: positive data points $\vect{y}=(y_1,y_2,\dots,y_N)^{\mathsf{T}}$, covariates $\vect{x}_1,\dots,\vect{x}_N$, and initial parameters $( \bfp , \bfT ,\theta)$}\\
\begin{enumerate} 
\item[ 1)]\textit{Transformation:} Transform the data into $z_i=g^{-1}(y_i;  \theta ) \exp(\vect{x}_i^\mathsf{T} \vect{\beta})$, $i=1,\dots,N$.

\item[ 2)]\textit{E-step:} compute the statistics (see Appendix \ref{apA} for precise definitions of the random variables $B_k$, $Z_k$, $N_{ks}$ and $N_k$)
\begin{align*}
    \mathbb{E}(B_k\mid \mat{Z}=\vect{z})=\sum_{i=1}^{N} \frac{\pi_k {\bfe_k}^{\mathsf{T}}\exp( \bfT z_i) \bft }{ \bfpi \exp( \bfT x_i) \bft }
\end{align*}
\begin{align*}
 \mathbb{E}(Z_k\mid \mat{Z}=\vect{z})=\sum_{i=1}^{N} \frac{\int_{0}^{z_i}{\bfe_k}^{\mathsf{T}}\exp( \bfT (z_i-u)) \bft  \bfpi \exp( \bfT u)\vect{e}_kdu}{ \bfpi \exp( \bfT z_i) \bft }   
\end{align*}
\begin{align*}
\mathbb{E}(N_{ks}\mid \mat{Z}=\vect{z})=\sum_{i=1}^{N}t_{ks} \frac{\int_{0}^{z_i}{\bfe_s}^{\mathsf{T}}\exp( \bfT (z_i-u)) \bft \, \bfpi \exp( \bfT u)\vect{e}_kdu}{ \bfpi \exp( \bfT z_i) \bft }
\end{align*}
\begin{align*}
\mathbb{E}(N_k\mid \mat{Z}=\vect{z})=\sum_{i=1}^{N} t_k\frac{ \bfpi  \exp( \bfT z_i){\bfe}y_k}{ \bfpi \exp( \bfT z_i) \bft }.
\end{align*}

\item[3)] \textit{M-step: let} 
\begin{align*}
\hat \pi_k=\frac{\mathbb{E}(B_k\mid \mat{Z}=\vect{z})}{N}, \quad \hat t_{ks}=\frac{\mathbb{E}(N_{ks}\mid \mat{Z}=\vect{z})}{\mathbb{E}(Z_{k}\mid \mat{Z}=\vect{z})}
\end{align*}
\begin{align*}
\hat t_{k}=\frac{\mathbb{E}(N_{k}\mid \mat{Z}=\vect{z})}{\mathbb{E}(Z_{k}\mid \mat{Z}=\vect{z})},\quad \hat t_{kk}=-\sum_{s\neq k} \hat t_{ks}-\hat t_k.
\end{align*}

\item[4)] Maximize
	\begin{align*}
	(\hat{ \theta },\hat{\vect{\beta}})  & = \argmax_{( \theta ,\vect{\beta})} \sum_{i=1}^N \log (f_{Y}(y_i; \hat{\bfp}, \hat{\bfT}, \theta ,\vect{\beta} ))
 \\
 & = \argmax_{( \theta ,\vect{\beta})} \sum_{i=1}^N \log\left(
 m(\vect{x}_i \vect{\beta} ) \lambda (y;  \theta )\, {\bfpi}\exp \left( m(\vect{x}_i \vect{\beta} )\int_0^y \lambda (s;  \theta )ds\ \mat{T} \right)\vect{t} 
 \right)
	\end{align*}

\item[5)] Update the current parameters to $({\bfp},\mat{T}, \theta ,\vect{\beta}) =(\hat{{\bfp}},\hat{\mat{T}}, \hat{ \theta },\hat{\vect{\beta}})$. Return to step 1 unless a stopping rule is satisfied.
\end{enumerate}
    \State \textit{\textbf{Output}: fitted representation $( \bfp , \bfT ,\theta, \vect{\beta})$.}
\end{algorithmic}
\end{algorithm}

A standard fact of the above procedure, which is not hard to verify directly, is the following result.

\begin{proposition}
The likelihood function is increasing at each iteration of Algorithm \ref{alg;IPHreg}. Thus, for fixed $p$, since in that case the likelihood is also bounded, we obtain convergence to a (possibly local) maximum.
\end{proposition}

 \begin{remark}[Computational remarks] \rm 
Algorithm \ref{alg;IPHreg} is simple to comprehend, and very often will converge not only to a maximum but to a global maximum. However, its implementation can quickly turn prohibitively slow if one fails to implement fast matrix analytic methods. The main difficulty is the computation of matrix exponentials, which appears in the E-step of step 1, and then repeatedly in the optimization of step 2.
This problem is by no means new, with \cite{moler1978nineteen} already providing several options on how to calculate the exponential of a matrix.

In \cite{asmussen1996fitting}, matrix exponentiation was done { for PH estimation} by converting the problem into a system of ordinary differential equations (ODE's), and then subsequently solved by the Runge-Kutta method of order four. The C implementation, called EMpht, is still available online, cf. \cite{olsson1998empht}. Using the same ODE approach, { \cite{bladt2021semi} implemented a C++ based algorithm} available from the \texttt{matrixdist} package in R (cf. \cite{matrixdist}). However, { when extended to IPH distributions,} this approach requires reorderings and lacks the necessary speed to estimate datasets of the magnitude of those found in insurance.

For this reason we presently choose to make use of the uniformization method, which consists on expressing the dynamics of a continuous-time Markov jump process in terms of a discrete-time chain where jumps occur at a Poisson intensity. {  More precisely, taking $\phi= \max_{k=1,\dots,p} (-t_{kk})$ and $\mat{Q} = {\phi}^{-1}\left(  \phi \mat{I} + \bfT \right)$, which is a transition matrix, we have that
\begin{align*}
 	\exp( \bfT y)= \sum_{n=0}^{\infty} \frac{\mat{Q} ^{n} (\phi y)^{n}}{n !} e^{-\phi y }  	\,. 
 \end{align*}
 and a truncated approximation of this series has error smaller than
  \begin{align*}
 	 \sum_{n=M+1}^{\infty} \frac{(\phi y)^{n}}{n !} e^{-\phi y } = \mathbb{P}(N_{\phi y}>M) \,,
 \end{align*}
 with $N_{\phi y} \sim \mbox{Poisson}(\phi y) $. In other words, with the introduction of the regularization parameter $\phi$, we are able to tame the error incurred in the truncation approximation by considering $M$ large enough according to a Poisson law.  Further improvements can be achieved by artifacts as the following one: since trivially $e^{\bfT y}=(e^{\bfT y/2^{m}})^{2^m}$, then the Poisson law with mean $\phi y /2^{m} <1 $ implies good approximations with small $M$ for $e^{\bfT y/2^{m}}$ and then $e^{\bfT y}$ is obtained by simple sequential squaring.

Integrals of matrix exponentials can be computed using an ingenious result of \cite{van1978computing}, which states that
	\begin{align*}
		\exp \left( \left(  \begin{array}{cc}
		\bfT & \bft \, \bfp \\
		\0 & \bfT
	\end{array}  \right) y \right)=  \left(  \begin{array}{cc}
		\ex^{\bfT y } & \int_{0}^{y} \ex^{ \bfT (y-u)} \bft \bfp \ex^{ \bfT u}du \\
		\0 & \ex^{\bfT y } 
	\end{array}  \right)\,.
	\end{align*}
	Hence, integrals of the form $\int_{0}^{y} \ex^{ \bfT (y-u)} \bft \bfp \ex^{ \bfT u}du$ can easily be computed at a stroke by one single matrix exponential evaluation of the left-hand-side expression and extracting the corresponding upper-right sub-matrix.
}

Bundled together with a new RcppArmadillo implementation (which is now also part of the latest Github version of \texttt{matrixdist}), the increase in speed was of about two to three orders of magnitude. Consequently, larger datasets with more covariates can be estimated in reasonable time. Naturally, the methods are still slower than those for simple regression models such as GLM's.

 \end{remark}

\subsection{Inference and goodness of fit.}
Inference and goodness of fit can always be done via parametric bootstrap methods. However, re-fitting a PH regression can be too costly. Consequently, we will take a more classical approach and derive certain quantities of interest which will allow us to perform variable selection and assess the quality of a fitted model.

\subsubsection{Inference for phase-type regression models}

For the following calculations, we focus only in the identifiable parameters of the regression, which are the relevant ones in terms of segmentation. Furthermore, we now make the assumption that $m$ is differentiable, and write $\ell_{\vect{y}}(\vect{\beta},\theta)$ for the joint log-likelihood function of the observed severities $\vect{y}=(y_1,\dots,y_N)$ with corresponding rating factors { $\overline{\vect{x}}=(\vect{x}_1,\dots,\vect{x}_N)$.

From \eqref{iph_density} we obtain for $j = 1,\dots,d$ and $h(y; \theta ):=\int_0^y \lambda (s;  \theta )ds$,
\begin{align}\label{deriv_loglik_betas}
\frac{d\ell_{\vect{y}}(\vect{\beta},\theta)}{d\beta_j}&= \sum_{i=1}^N G_1(i,j|\bfp,\bfT,\vect{\beta},\theta,\vect{y},\overline{\vect{x}}),\\
\frac{d\ell_{\vect{y}}(\vect{\beta},\theta)}{d \theta }&= \sum_{i=1}^N G_2(i|\bfp,\bfT,\vect{\beta},\theta,\vect{y},\overline{\vect{x}}),\label{deriv_loglik_theta}
\end{align}

where
\begin{align*}
G_1(i,j|\bfp,\bfT,\vect{\beta},\theta,\vect{y},\overline{\vect{x}})=x_{ij}m'(\vect{x}_i^\mathsf{T}\vect{\beta} ) \left(\frac{1}{m(\vect{x}_i^\mathsf{T}\vect{\beta} ) }+
\frac{{\bfpi}\exp \left( m(\vect{x}_i^\mathsf{T}\vect{\beta} )h(y; \theta )\mat{T} \right)h(y_i; \theta )\mat{T}\vect{t}}{{\bfpi}\exp \left( m(\vect{x}_i^\mathsf{T}\vect{\beta} )h(y_i; \theta )\ \mat{T} \right)\vect{t}}
\right),\\
G_2(i|\bfp,\bfT,\vect{\beta},\theta,\vect{y},\overline{\vect{x}})=\frac{\frac{d}{d \theta }\lambda(y_i; \theta )}{\lambda(y_i; \theta )}+
\frac{{\bfpi}\exp \left( m(\vect{x}_i^\mathsf{T}\vect{\beta} )h(y_i, \theta ) \mat{T} \right)m(\vect{x}_i^\mathsf{T}\vect{\beta})\frac{d}{d \theta }h(y_i, \theta )\mat{T}\vect{t}}{{\bfpi}\exp \left( m(\vect{x}_i^\mathsf{T}\vect{\beta} )h(y_i; \theta ) \mat{T} \right)\vect{t}}
\end{align*}

Such expressions should be equal to zero whenever convergence is achieved in step 2 of Algorithm \ref{alg;IPHreg}.

In general, the score functions \eqref{deriv_loglik_betas} and \eqref{deriv_loglik_theta} will not have an explicit solution when equated to zero. Concerning the $(d+1)\times(d+1)$-dimensional Fisher Information matrix, we may write it as
\begin{align}\label{fisherinfo}
[\mathcal{I}]_{jk}&= \begin{cases}
 \sum_{i=1}^N G_1(i,j|\bfp,\bfT,\vect{\beta},\theta,\vect{y},\overline{\vect{x}}) G_1(i,k|\bfp,\bfT,\vect{\beta},\theta,\vect{y},\overline{\vect{x}}) & 1\le j,k \le d,\\
  \sum_{i=1}^N G_1(i,j|\bfp,\bfT,\vect{\beta},\theta,\vect{y},\overline{\vect{x}}) G_2(i|\bfp,\bfT,\vect{\beta},\theta,\vect{y},\overline{\vect{x}}) & 1\le j \le d,\;k=d+1,\\
   \sum_{i=1}^N G_2^2(i|\bfp,\bfT,\vect{\beta},\theta,\vect{y},\overline{\vect{x}}) & j=k=d+1.
\end{cases}
\end{align}
Then we may use the asymptotic MLE approximation, as $N\to\infty$, 
$$\widehat{(\vect{\beta},\theta)}\stackrel{d}{\approx}\mathcal{N}((\vect{\beta},\theta),\mathcal{I}^{-1}),$$ which can be used to compute standard errors and Neyman-type confidence intervals. For instance, $$(\hat\beta_1-1.96 \sqrt{[\mathcal{I}^{-1}]_{11}},\:\hat\beta_1+1.96 \sqrt{[\mathcal{I}^{-1}]_{11}}),$$ constitutes an approximate $95\%$ confidence interval for the parameter $\beta_1$. We obtain p-values in much the same manner.
}

Observe that we have deliberately omitted inference for the PH parameters $\bfp$ and $\bfT$. The reason being, this is a particularly difficult task due to the \textit{non-identifiability} issue of PH distributions, namely that several representations can result in the same model. Papers such as \cite{bladt2011fisher, zhang2021efficient} provide methods of recovering the information matrix for these parameters, but the validity of the conclusions drawn from such approach are not fully understood. Consequently, and similarly to other regression models, we will only perform estimation on the regression coefficients themselves, and use the distributional parameters merely as a vehicle to obtain $\vect{\beta}$.

\begin{remark}\rm
We performed estimation using numerical optimization, with and without using the gradient of the log-likelihood for step 2 in Algorithm \ref{alg;IPHreg}. Not using the gradient was always faster, since when dealing with matrix exponentials the evaluation of the derivatives is equally costly as the objective function itself, and thus best avoided.
{ In practice, the Fisher Information matrix may be near-singular when large numbers of covariates are considered. The negative of the the Hessian matrix obtained by numerical methods such as BFGS is a more reliable alternative in this case, whose inverse may also be used to approximate the asymptotic covariance matrix.}
\end{remark}

\subsubsection{Goodness of fit for phase-type regression models}

Below is a { brief description of a goodness of fit diagnostic tool for the above models, which, when ordered and plotted against theoretical uniform order statistics, constitute a PP-plot}. The idea is to transform the covariate-specific data into a parameter-free scale using the probability integral transformation (PIT).

{ Indeed, by} applying the PIT transform to the data
{ 
\begin{align}\label{pp_specification}
r_i=\left({\bfpi}\exp \left(m(\vect{x}_i^\mathsf{T} \vect{\beta} ) \int_0^{y_i} \lambda (s; \theta )ds\ \mat{T} \right)\vect{e}\right),\quad i=1,\dots N.
\end{align}}
The null hypothesis under the PH regression model is $$\mathcal{H}_0:Y_i\sim \mbox{IPH}({\bfp},\mat{T},m(\vect{x}_i^\mathsf{T}\vect{\beta})\lambda),\quad i=1,\dots,N,$$
and it { standard} see that the dataset 
\begin{align*}
\mathcal{D}=\{r_1,\:r_2,\dots, r_N\}
\end{align*}
constitutes a sample of { uniform} variables.

Thus, it only remains to assess the goodness of fit of $\mathcal{D}$ against a standard { uniform} distribution, for which a suite of statistical tests and visual tools can be applied.

\subsection{On the choice of matrix dimension and structure}
Often a general sub-intensity matrix structure is too general to be either practical or parsimonious, and special structures can do the job almost as well with fewer parameters (see for instance \cite{Bladt2017}, Section 3.1.5, and also Section 8.3.2). This is because IPH distributions are not identifiable: two matrices of different dimension and/or structure may yield exactly the same distribution.

A zero in the off-diagonal of the sub-intensity matrix $\bfT$ indicates absence of jumps between two states, whereas a zero in the diagonal is not possible. Below we describe the most commonly used special structures for $\bfT$ (and its corresponding $\bfp$). Note that we borrow the names of the structures from the homogeneous case, but the resulting distribution will in general not be linked to such name. For instance, a Matrix-Pareto of exponential structure will yield a Pareto distribution, and not an exponential one.

\begin{figure}[!htbp]
        \centering
        \begin{subfigure}[b]{0.49\textwidth}
            \centering
\includegraphics[clip, trim=16cm 19cm 13cm 15cm,width=\textwidth]{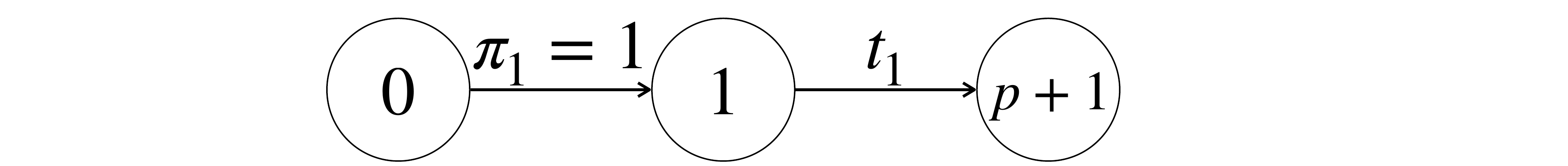}
            \caption[Exponential]%
            {{\small Exponential}}    
            \label{fig:Exponential_structure}
        \end{subfigure}
        \hfill
        \begin{subfigure}[b]{0.49\textwidth}  
            \centering 
\includegraphics[clip, trim=16cm 19cm 13cm 15cm,width=\textwidth]{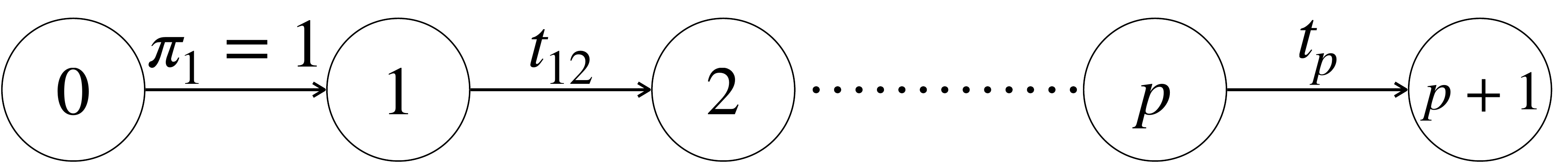}            \caption[Erlang]%
            {{\small Erlang}}    
            \label{fig:Erlang_structure}
        \end{subfigure}
        \vskip\baselineskip
        \begin{subfigure}[b]{0.49\textwidth}   
            \centering 
\includegraphics[clip, trim=16cm 11cm 13cm 7cm,width=\textwidth]{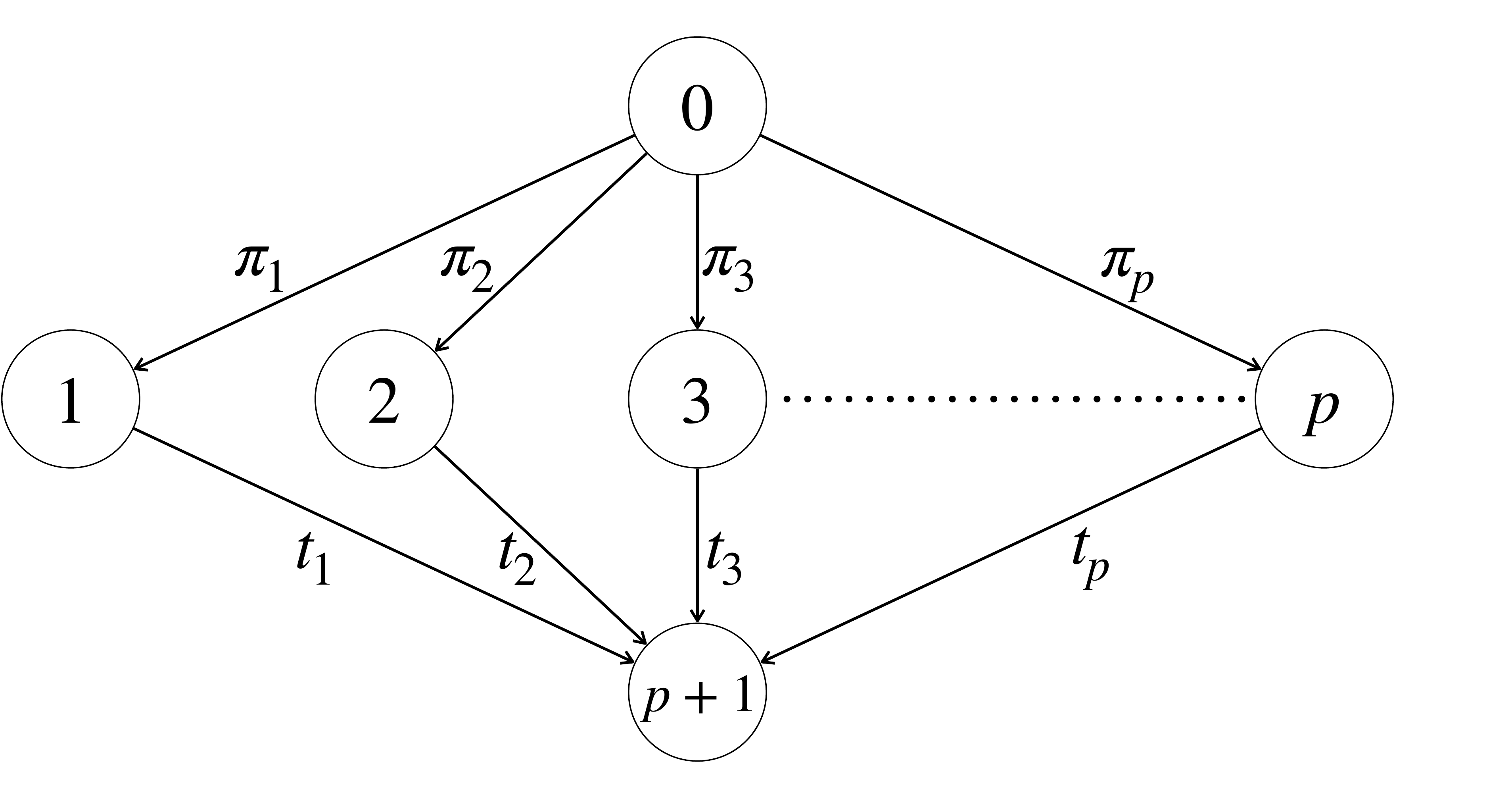}            \caption[Hyper-exponential]%
            {{\small Hyper-exponential}}    
            \label{fig:Hyper-exponential_structure}
        \end{subfigure}
        \hfill
        \begin{subfigure}[b]{0.49\textwidth}   
            \centering 
\includegraphics[clip, trim=15cm 6.5cm 15cm 15cm,width=\textwidth]{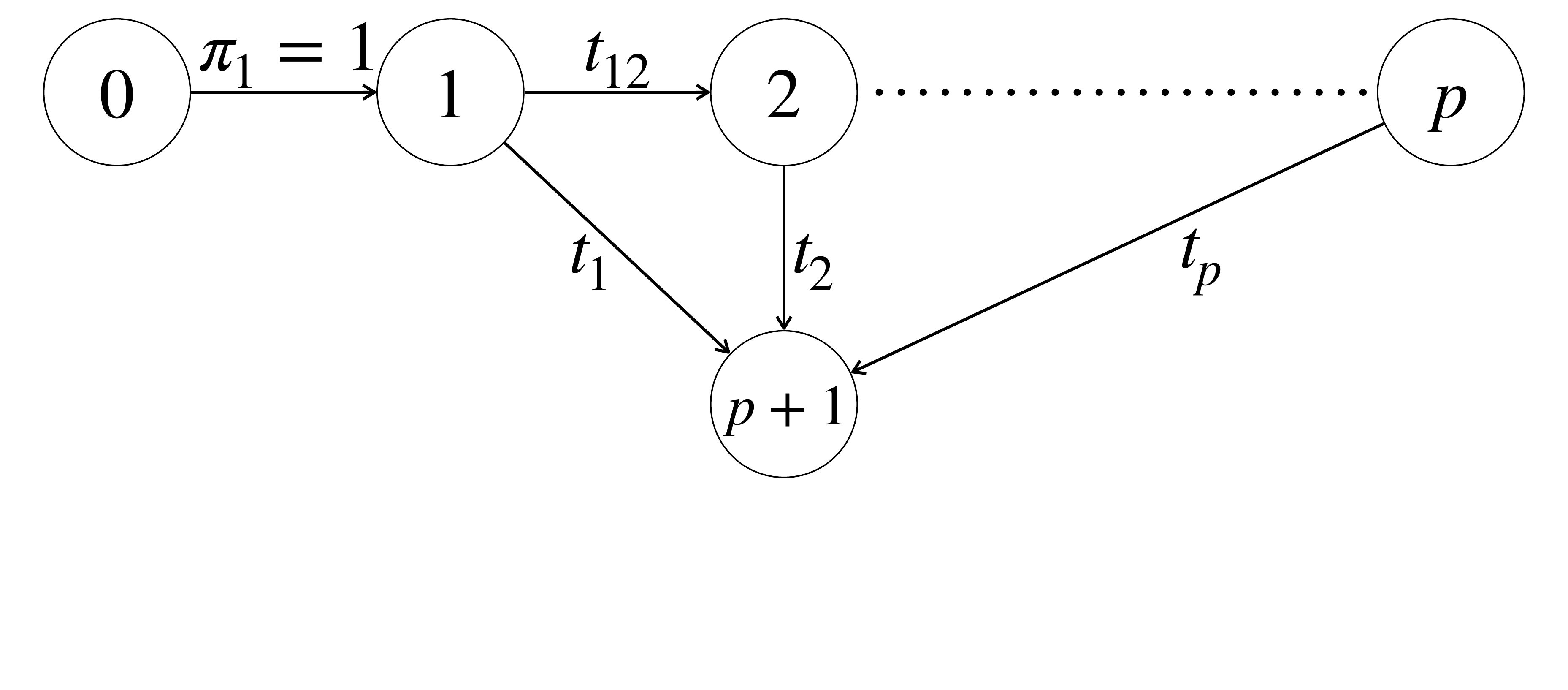}            \caption[Coxian]%
            {{\small Coxian}}    
            \label{fig:Coxian_structure}
        \end{subfigure}
                \vskip\baselineskip
        \begin{subfigure}[b]{0.49\textwidth}   
            \centering 
\includegraphics[clip, trim=15cm 6.5cm 15cm 7cm,width=\textwidth]{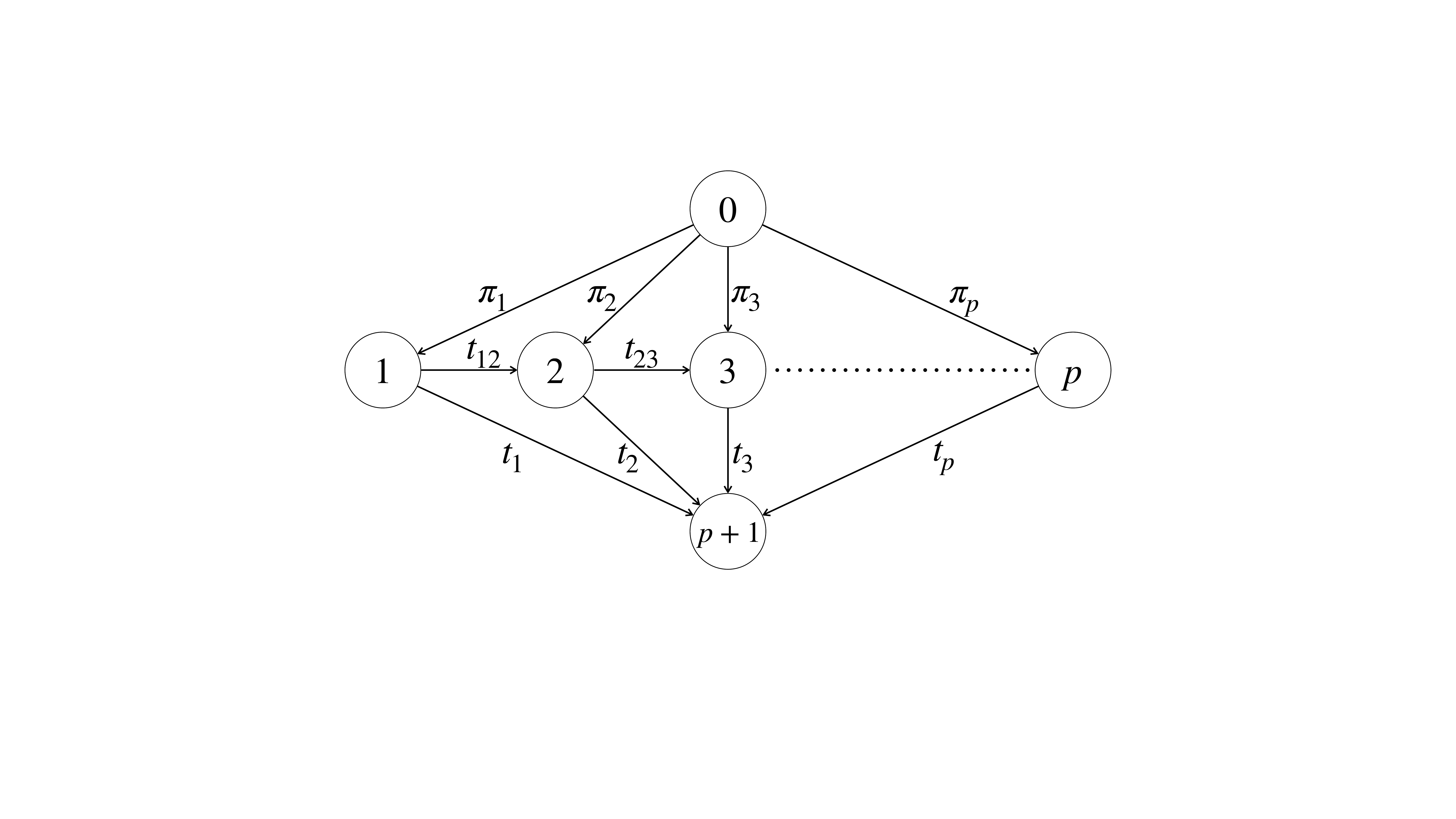}           \caption[generalized Coxian]%
            {{\small generalized Coxian}}    
            \label{fig:generalized_Coxian_structure}
        \end{subfigure}
        \hfill
        \begin{subfigure}[b]{0.49\textwidth}   
            \centering 
\includegraphics[clip, trim=15cm 3cm 15cm 7cm,width=\textwidth]{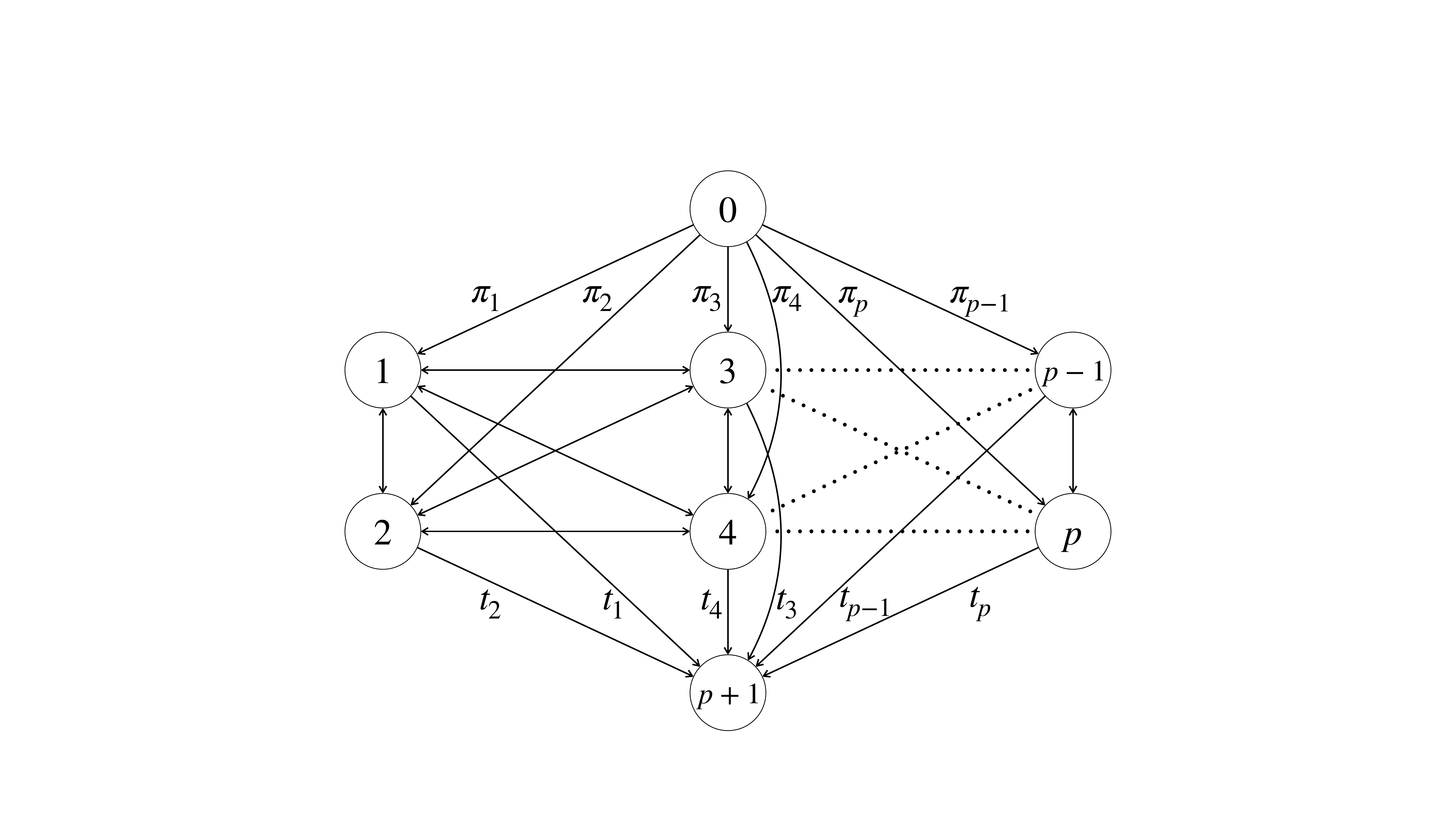}        \caption[General]%
            {{\small General}}    
            \label{fig:General_structure}
        \end{subfigure}
        \caption[ Underlying Markov structures. ]
        {\small Underlying Markov structures. Names are borrowed from the corresponding PH representations, but apply to our inhomogeneous setup as well. The state $0$ is added for schematic reasons, but is not part of the actual state-space of the chain. The (F) General case has the intensities $t_{ij}$ and $t_{ji}$ between each pair of states $i,j\in\{0,\dots,p\}$ omitted for display purposes.} 
        \label{fig:structures}
\end{figure}

\textit{Exponential structure.}
This is the simplest structure one can think of, and applies only for $p=1$, i.e. the scalar case. We have $\bfpi=1$, $\bfT=-\alpha<0$ and so $\bft=\alpha$. The corresponding stochastic process is schematically depicted in the top-left panel of Figure \ref{fig:structures}.

\textit{Erlang structure.}
This structure traverses $p$ identical states consecutively from beginning to end, or in matrix notation:
\[ \bfpi =(1, \:0,  \dots, \:0 ),\]
\[  \bfT = \begin{pmatrix}
        -\alpha & \alpha &  & 0 \\
         &   \ddots&\ddots  & \vdots\\
          &   &  -\alpha& \alpha\\
        0 &  & &-\alpha \end{pmatrix}, \, \quad \bft= \begin{pmatrix}0 \\ \vdots \\ 0 \\\alpha \end{pmatrix},\quad \alpha>0. \]
The corresponding stochastic process is schematically depicted in the top-right panel of Figure \ref{fig:structures}.

\textit{Hyperexponential PH distribution.}
This structure will always give rise to a mixture of scalar distributions, since it starts in one of the $p$ states but has no transitions between them thereafter. The matrix representation is the following:
\[ \bfpi =(\pi_1, \: \pi_2,  \dots, \: \pi_p),\]
\[  \bfT = \begin{pmatrix}
        -\alpha_1 &  &  & 0 \\
         &   & \ddots & \\
         0& & & -\alpha_p \end{pmatrix}, \, \quad \bft = \begin{pmatrix}\alpha_1 \\ \vdots  \\\alpha_p \end{pmatrix},\quad \alpha_i>0, \:\:\: i=1,\dots,p. \]
The corresponding stochastic process is schematically depicted in the center-left panel of Figure \ref{fig:structures}.

\textit{Coxian structure.}
This structure is a generalization of the Erlang one, where different parameters are allowed for the diagonal entries, and also absorption can happen spontaneously from any intermediate state. The parametrisation is as follows:
\[ \bfpi =(1,\:  0,  \dots,\: 0),\]
\[  \bfT = \begin{pmatrix}
        -\alpha_1 & q_1\alpha_1 &  & 0 \\
        & -\alpha_2 & q_2\alpha_2& \\
          &  & \ddots \ddots&  \\
          &  &  -\alpha_{p-1}& q_{p-1}\alpha_{p-1} \\
        0 &  & & -\alpha_p \end{pmatrix}, \quad \bft = \begin{pmatrix}(1-q_1)\alpha_1 \\ (1-q_{2})\alpha_{2} \\ \vdots \\ (1-q_{p-1})\alpha_{p-1} \\\alpha_p \end{pmatrix},\]
with $\alpha_i>0,\:q_i\in[0,1], \:i=1,\dots,p$.
The corresponding stochastic process is schematically depicted in the centre-right panel of Figure \ref{fig:structures}.

\textit{Generalized Coxian structure.}
This structure has the same $\bfT$ and $\bft$ parameters as in the Coxian case, the only difference being that the initial vector is now distributed among all $p$ states, that is:
\[ \bfpi =(\pi_1, \: \pi_2,\dots, \:\pi_p ).\]
The corresponding stochastic process is schematically depicted in the bottom-left panel of Figure \ref{fig:structures}.

\begin{remark}\rm
One convenient feature of having zeros in a sub-intensity matrix is that the EM-algorithm will forever keep those zeros in place, effectively searching only within the sub-class of IPH distributions of a given special structure. A prevalent observation which has nearly become a rule of thumb is that Coxian structures perform nearly as well as general structures, and are certainly much faster to estimate.
\end{remark}

\begin{remark}\rm
{ When dealing with matrices whose entries are the parameters of the model, the usual systematic measures such as AIC or BIC are overly conservative (yielding too simple models), and the appropriate amount of penalization is a famous unsolved problem in the PH community, and out of the scope of the current work. The selection of the number of phases and the special sub-structure of such matrix is commonly done by trial and error, much in the same manner as is done for the tuning of  hyperparameters of a machine learning method, or the selection of the correct combination of covariates in a linear regression. Such approach is also taken presently. Despite this disadvantage, there is one advantage from a purely statistical standpoint, which is that standard errors (and thus significance of regression coefficients) arising from asymptotic theory are more truthful if no automatic selection procedure is used.

Hence, when selecting between fitted dimensions and structures it is advised to keep track of the negative log-likelihood (rather than AIC or BIC) and its decrease when increasing the number of phases or changing the structure. Theoretically, larger matrices will always yield better likelihood, but in practice it is often the case that the likelihood stops increasing (effectively getting stuck in a good local optimum), or increases are not so significant. Usually, EM algorithms for PH distributions for marginal distributions estimate effectively and in reasonable time up to about 20-30 phases, while for a phase-type regression the number is closer to 5-10. The additional inhomogeneity function of IPH distributions, however, makes models more parsimonious than the PH counterparts, and usually less than 10 phases are needed.

Once the dimension and structure is chosen, one can perform model selection with respect to the regression coefficients. These fall into the usual framework and the AIC or BIC criteria can then be used to compare and select between various models.
}
\end{remark}

\section{A simulation study}\label{sec:sim}
Before applying the above models to a real-life insurance dataset, we would first like to confirm that PH regression is an effective tool for estimating data exhibiting features which are common in insurance. Namely, claim severity can exhibit multimodality and different behaviour in the body and tails of the distribution, the latter being heavy-tailed in some lines of business. In order to create synthetic data with these features we take the following approach. 

{ We simulate $N=1000$ times from a bivariate Gaussian copula $\bfX=(X_1,X_2)^\mathsf{T}$ with correlation coefficient $0.7$, which will play the role of sampling from a bivariate feature vector with uniform marginals.} Then we create three mean specifications based on the first entry of such vector:
\begin{align*}
\mu_1=\exp(X_1),\quad \mu_2=\exp(3+X_1), \quad \mu_3=\exp(X_1-1),
\end{align*}
in a sense individually specifying a log-link function. We finally sample with probabilities $0.4,\,0.4,\,0.2$ from three different regression specifications:
\begin{align*}
\text{Gamma}(\mu=\mu_1, \phi=1),\quad \text{Gamma}(\mu=\mu_2, \phi=1),\quad \exp(\text{Gamma}(\mu=\mu_3, \phi=1)).
\end{align*}

\begin{figure}[!htbp]
\centering
\includegraphics[width=0.7\textwidth]{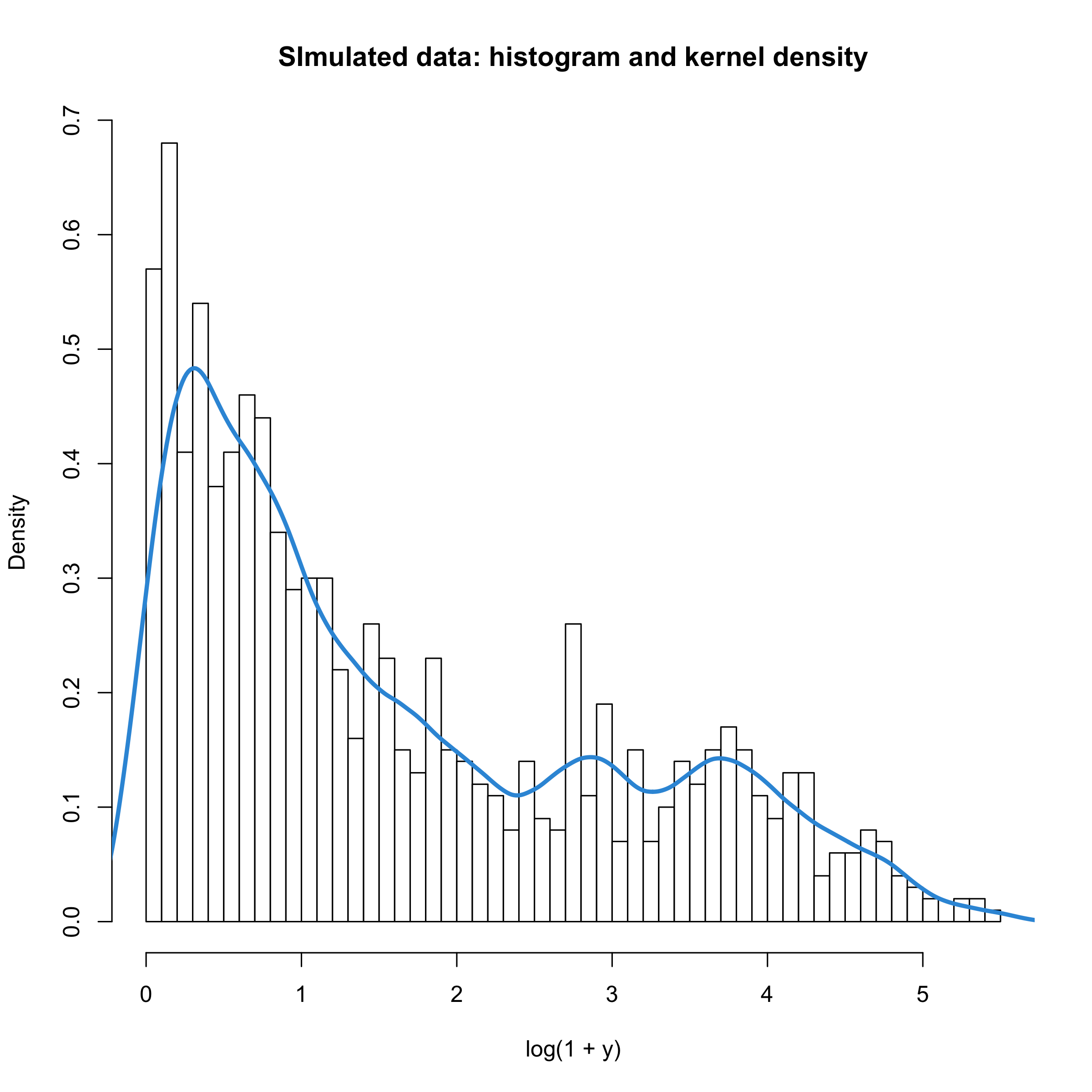}
\caption{Histogram and kernel density estimate of the log-transformed simulated data.
} \label{sim_data_dens}
\end{figure}

The data is depicted in Figure \ref{sim_data_dens}.

Observe that the third component implies Pareto-type tails, although this component has the smallest probability of occurrence. We also remark that the true driver of the regression is $X_1$, and thus we would ideally like to obtain the non-significance of $X_2$ from the inference.

\begin{figure}[!htbp]
\centering
\includegraphics[width=0.7\textwidth]{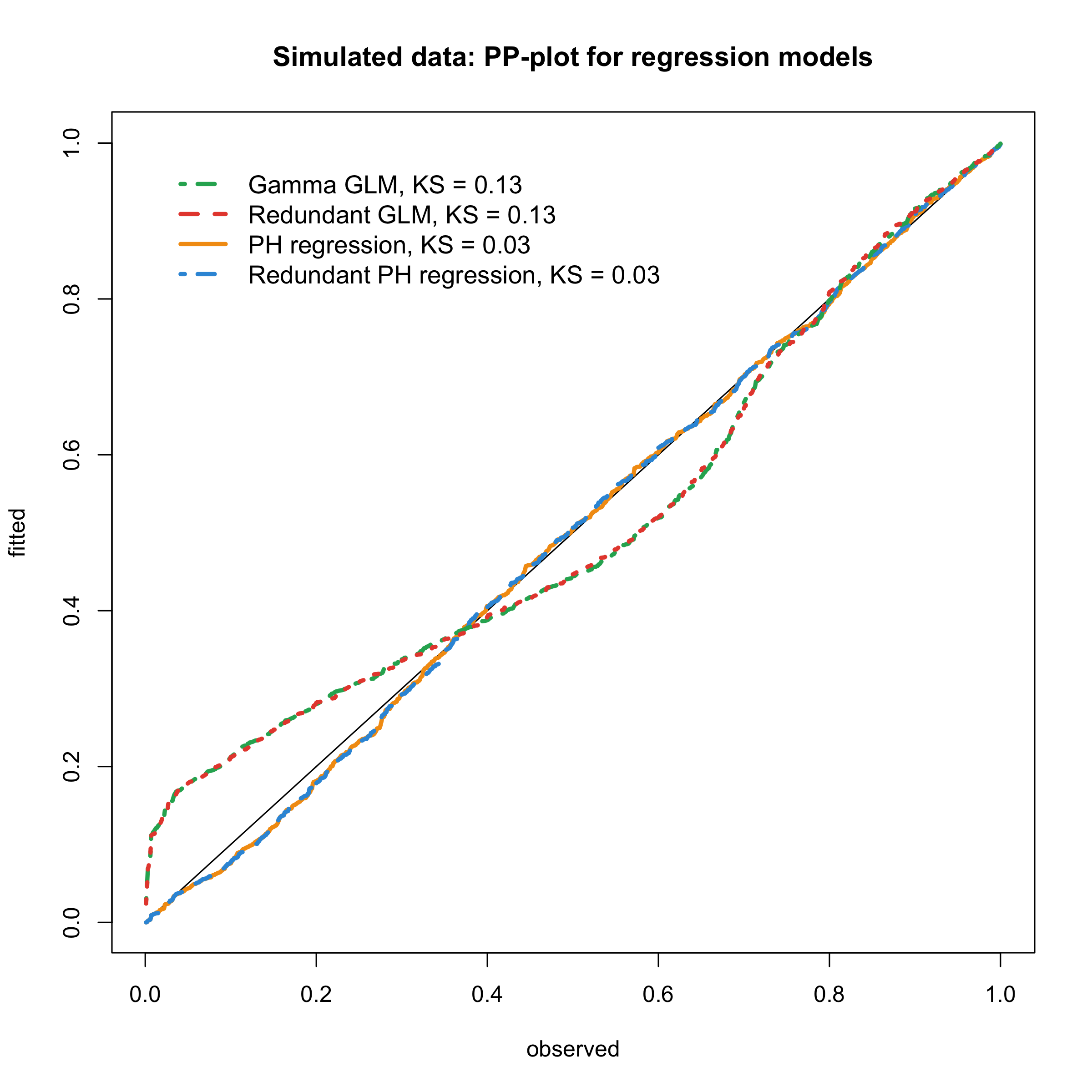}
\caption{ {  Ordered PIT's from equation \eqref{pp_specification} versus uniform order statistics for the simulated dataset. {  KS refers to the Kolmogorv-Smirnov statistic for testing uniformity}.}
} \label{sim_reg}
\end{figure}

Next, consider four model specifications: 
\begin{enumerate}
\itemsep0.8em 
\item a Gamma GLM with log-link function and considering $X_1$ as covariate. 
\item a Gamma GLM with log-link function and considering $X_1$ and $X_2$ as covariates. 
\item a Matrix-Pareto PH regression of Coxian type and dimension $3$, considering $X_1$ as covariate. 
\item a Matrix-Pareto PH regression of Coxian type and dimension $3$, considering $X_1$ and $X_2$ as covariates.
\end{enumerate}
The analysis was carried using the \texttt{matrixdist} package in R, \cite{matrixdist}. The results are given in Table \ref{table:coefficients0} (observe that the IPH models do not have intercept since it is included in their sub-intensity matrices), showing good segmentation for the matrix-based methods. Uninformative covariates can play the role of mixing in the absence of a correctly specified probabilistic model, and thus we see that $X_2$ is significant for the GLM model, whereas this is no longer the case for the PH regression. The AIC and BIC select the overall best model to be the Matrix-Pareto with only $X_1$ as rating factor. {  In terms of goodness-of-fit, we visually confirm from Figure \ref{sim_reg} that the distributional features of the data are captured substantially better by considering an underlying Markov structure.}

\begin{table}[!htbp]
\begin{center}
\caption{GLMs and PH regression models}

\begin{tabular}{l c c c c}
\hline
& Gamma GLM & Gamma GLM & M-Pareto(3) & M-Pareto(3) \\
\hline
Intercept    & $1.936^{***}$ &$2.054^{***}$ &  &        \\
               & $(0.127)$    &  (0.121)  &         &                 \\
$X_1$      & $0.762^{***}$ &$1.138^{***}$ & $-0.813^{***}$ &  $-1.039^{***}$      \\
               & $(0.280)$    &  (0.2105)  & $(0.187)$        &        $(0.147)$           \\
$X_2$        & $-0.586^{*}$ &  & $ -0.266^{}$   &                   \\
               & $(0.273)$    &    & $(0.181)$        &                   \\
\hline
AIC            & $6,407$     & $6,411$   & $6,100$       & $\textbf{6,098}$       \\
BIC            & $6,426$    & $6,426$    & $6,139$       & $\textbf{6,133}$       \\
Log Likelihood & $-3,199$     & $-3,202$   & $-3,042$       & $-3,042$       \\
Degrees of freedom       & $4$  & $3$ & $8$ & $7$ \\
Num. obs.      & $1,000$      & $1,000$       & $1,000$            & $1,000$            \\
\hline
\multicolumn{4}{l}{\scriptsize{$^{***}p<0.001$; $^{**}p<0.01$; $^{*}p<0.05$}}
\end{tabular}
\label{table:coefficients0}
\end{center}
\end{table}

\section{Application to a the French Motor Personal Line dataset}\label{sec:reg}

In this section, we consider a real-life insurance dataset: the publicly available French Motor Personal Line dataset. We apply {  the IPH and PH regression estimation procedures to the distribution of claim severities, illustrating how the theory and algorithms from the previous sections} can be effectively applied to model claim sizes.

We consider jointly the four datasets \texttt{freMPL1},  \texttt{freMPL2}, \texttt{freMPL3}, \texttt{freMPL4}, from the \texttt{CASdatasets} package in  \texttt{R}. These data describe claim frequencies and severities pertaining to four different coverages in a French motor personal line insurance for about 30,000 policies in the year 2004. { The data has 18 covariates, which are described in the supplementary material. The $7008$ observations with positive claim severity possess a multimodal density, arising from mixing different types of claims.}

\subsection{Marginal estimation}
{  In a first step, we estimate the marginal behavior of the data, without rating factors.} The model we choose to estimate is an IPH distribution with the exponential transform, namely, a Matrix-Pareto distribution. This choice is motivated by a preliminary analysis of the data by which a Hill plot indicated the presence of regularly-varying tails, indicating heavy-tailedness. Since the data on the log-scale has a pronounced hump in the middle, a dimension larger than one or two is needed, so we chose five. Finally, since there are no particular specificities in the right or left tail, a Coxian sub-matrix structure can do an equally good job as a general structure, and thus we select the former in the name of parsimony. 

We also estimate a $20$-dimensional Matrix-Weibull distribution on the data { to illustrate the denseness property of IPH distributions}. The Matrix-Weibull distribution generates modes more easily than the Matrix-Pareto, and we choose such a high dimension in order to illustrate the denseness of IPH distributions. However, this model is over-parametrized and thus mainly of academic interest.

As a first reference, we also fit a two-parameter Gamma distribution to the data. We also consider a splicing (or composite) model, cf. \cite{albrecher2017reinsurance,reynkens2017modelling} (see also \cite{miljkovic2016modeling}), defined as follows. Let $f_1,\:f_2$ be two densities supported on the positive real line. A spliced density based on the the latter components is then given by
\begin{align*}
f_S(x)=v \frac{f_1(x)}{F_1(t)} 1_{(0,t]}(x) + (1-v) \frac{f_2(x)}{1- F_2(t)} 1_{(t,\infty)}(x), \quad t>0,\:\: v\in[0,1],
\end{align*}
where $t$ is the splicing location. We consider a popular implementation found in the \texttt{ReIns} package in R, \cite{reins}, where $$f_1(x)=\sum_{j=1}^{M} \omega_{j} \frac{x^{r_{j}-1} \exp (-x / \zeta)}{\zeta^{r_{j}}\left(r_{j}-1\right) !},\:\:x>0,\quad f_2(x)=\frac{1}{\nu t}\left(\frac{x}{t}\right)^{-\frac{1}{\nu}-1},\:\:x>t,$$ are, respectively, a mixture of Erlang (ME) densities with common scale parameter and a Pareto density, which we call the ME-Pareto specification.

The parameter estimates are the following for the Gamma distribution, found by MLE,
$$\text{Gamma shape}= 0.75, \quad \text{Gamma scale} = 2925.67,$$

while for the splicing model we obtain, selecting $\nu$ through EVT techniques (Hill estimator) and the remaining parameters through MLE (using an EM algorithm), and choosing $M$ from $1$ to $10$ according to the best AIC:
\begin{align*}
&M=5, \quad \zeta=363.7,\quad \vect{r}=(1,\,  2 ,\, 4,\, 11,\, 21),\\
 &\vect{\omega}=(0.26,\, 0.07,\, 0.53,\, 0.09,\, 0.02),\quad \nu =0.56,\quad v=0.96,
\end{align*}
and $t\approx10000$ was selected by visual inspection according to when the Hill plot stabilises for the sample.
%
Finally, for the IPH Matrix-Pareto distribution we obtain the following estimates by applying Algorithm \ref{alg;IPHreg} { without covariates} (which is implemented in the \texttt{matrixdist} package in R, \cite{matrixdist}):
$$\bfp=\bfe_1, \quad \bfT= \begin{pmatrix}
-12.61 & 12.48 & 0 & 0 & 0\\
0 & -12.61 & 10.33 & 0 & 0\\
0 & 0 & -1.99 & 1.99 & 0\\
0 & 0 & 0 & -7.34 & 7.34\\
0 & 0 & 0 & 0 & -7.34\\
\end{pmatrix},\quad \eta = 1149.57,$$
where $\bfe_1=(1,\,0,\,0,\,0,\,0)^\mathsf{T}.$ The $404$ fitted parameters of the Matrix-Weibull, most of the zero, are omitted, since they do not provide further intuition on IPH fitting.

Table \ref{table:marginal_fits} provides numerical results on the three fitted models, and in particular it shows that an IPH Matrix-Pareto distribution is well-aligned with the state-of-the-art models for severity modeling in insurance such as composite methods. The AIC and BIC are not good measures for IPH distributions since they can over-penalise them when the underlying sub-intensity matrix is not of minimal order. However, we see that even with this hinderance, the performance in terms of information criteria is positive.

\begin{table}[!htbp]
\begin{center}
\caption{Summary of fitted marginal models to severities from the freMPL dataset.}
\begin{tabular}{l c c c c}
\hline
& Gamma & Spliced ME-Pareto & Matrix-Pareto & Matrix-Weibull\\
\hline
Log Likelihood & $-60,653$     & $-59,611$   & $-59,605$   & $-59,167$    \\
Degrees of freedom       & $2$   & $12$ & $10$ & $404$\\
AIC            & $121,311$     & $119,247$   & $119,231$ & $\textbf{119,142}$          \\
BIC            & $121,325$    & $119,330$    & $\textbf{119,299}$   & $121,912$        \\
Num. obs.      & $7,008$      & $7,008$       & $7,008$    & $7,008$         \\
\hline
\end{tabular}
\label{table:marginal_fits}
\end{center}
\end{table}

\begin{figure}[!htbp]
\centering
\includegraphics[width=0.49\textwidth]{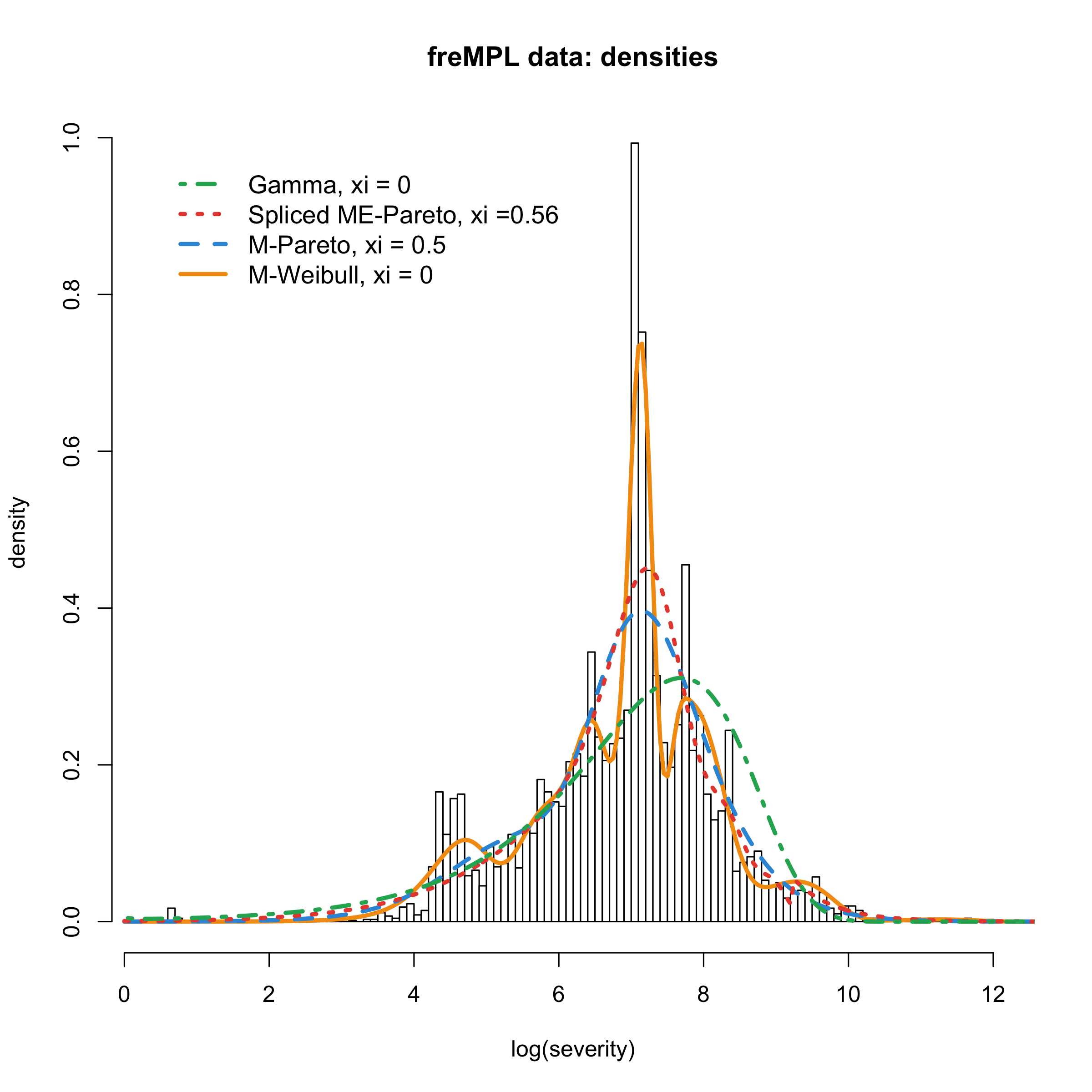}
\includegraphics[width=0.49\textwidth]{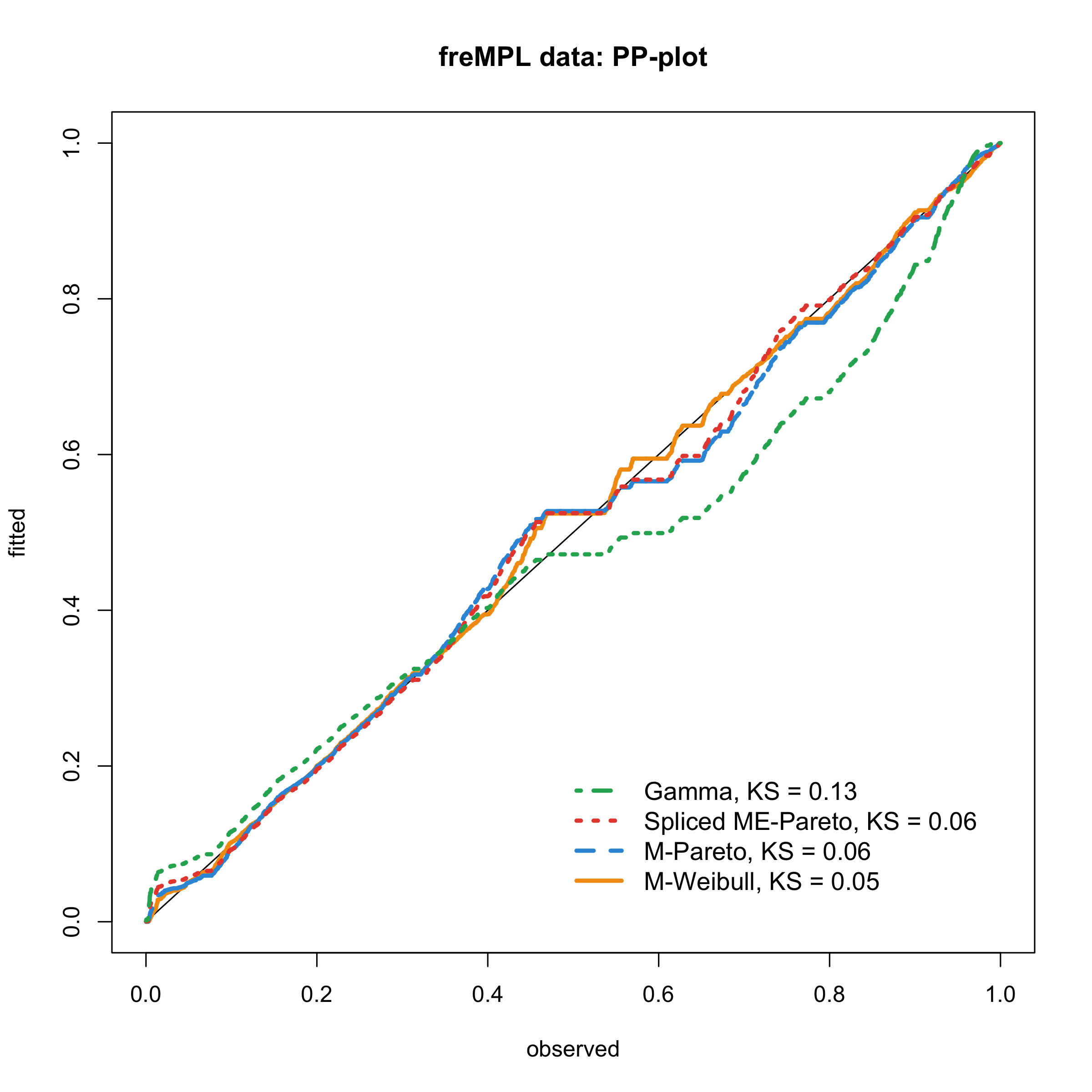}
\includegraphics[width=0.51\textwidth]{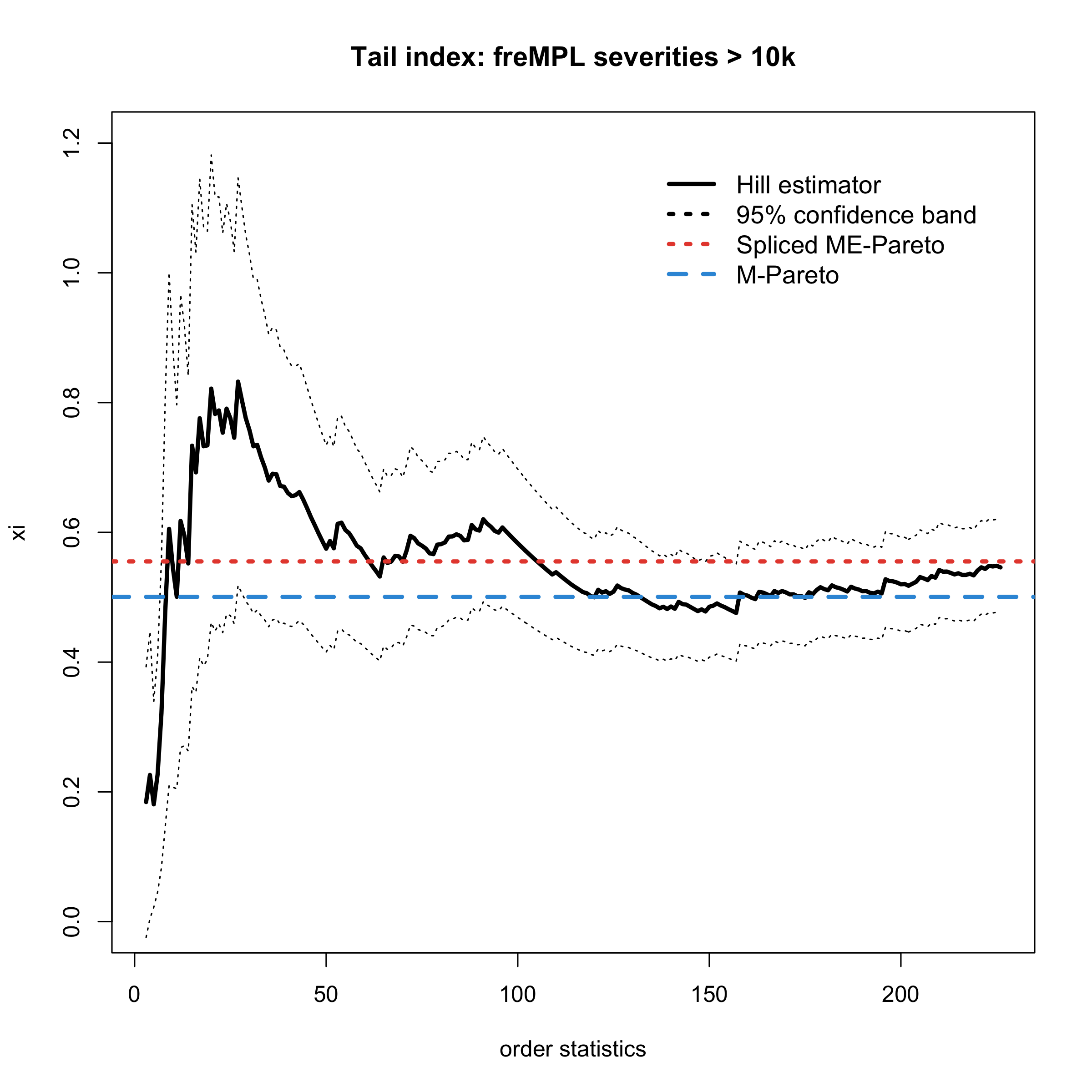}
\caption{ Fitted distributions to claim severities from the MPL dataset. {  The log-transform is used exclusively for visual purposes, the estimation having been carried out in the usual scale.}  {  KS refers to the Kolmogorv-Smirnov statistic for testing uniformity}.
} \label{freMPL}
\end{figure}

The quality of the fit is also assessed in Figure \ref{freMPL}. In particular, we observe that the data has a big spike at the log-severity around the value $7.3$, which poses problems to the first three distributions. This disturbance manifests itself in the PP-plot as a slight ``S"-shape. Not surprisingly, the visual quality of the fit of the splicing and IPH models is, as expected, far superior to its Gamma counterpart, while the difference between the two Pareto-tailed models is not very noticeable. Their implied tail indices agree, as given respectively by the formulas $\hat\xi=\nu=0.56,$ and
$$\hat\xi = -1/\max\{\Re\text{Eigen}(\hat\bfT)\} = 0.50,$$
where $\Re\text{Eigen}(\mat{A})$ denotes the set of real eigenvalues of a matrix $\mat{A}$. For reference, recall that in terms of the Pareto tail parameter, the following relationship holds $\alpha = 1/\xi$. As a sanity check, we see that both estimates for the tail index
 fall perfectly within the Hill estimator's bounds. In particular, the estimated models have finite mean (their tail index is less than one; alternatively, their Pareto index is larger than one). We would like to remark that fitting a conventional Pareto distribution as a global model will result in a particularly bad fit, such that in the IPH case we have managed use matrix parameters to our advantage, in order to automatically engineer an all-inclusive model, featuring no threshold selection.
 

\subsection{Incorporating rating factors}

Having already analysed the marginal behaviour of the French MPL data, we proceed to incorporate rating factors through the PH regression model that we have presented above. In Table \ref{severity_summary} we show summarising statistics for claim severity, while two further tables of summarising statistics for continuous rating factors and categorical rating factors, respectively, can be found in supplementary material. 

\begin{table}[!htbp] \centering 
  \caption{Summary statistics of the French MPL dataset: \textbf{claim severity}} 
  \label{severity_summary} 
\begin{tabular}{@{\extracolsep{5pt}}lccccccc} 
\hline \\[-1.8ex] 
Variable & \multicolumn{1}{c}{Mean} & \multicolumn{1}{c}{St. Dev.} & \multicolumn{1}{c}{Min} & \multicolumn{1}{c}{Pctl(25)} & \multicolumn{1}{c}{Median} & \multicolumn{1}{c}{Pctl(75)} & \multicolumn{1}{c}{Max} \\ 
\hline \\[-1.8ex] 
ClaimAmount & 2,176.9 & 5,807.2 & 0.4 & 500.4 & 1,204 & 2,163.2 & 163,427 \\ 
\hline \\[-1.8ex] 
\end{tabular} 
\end{table} 

To reduce the large number of categories and variables, we perform a pre-processing step where we individually, for each categorical variable, prune a regression tree according to the best predictive performance on the mean of the log-severities (since the variables exhibit heavy-tails). We then merge categories together according to the resulting optimal tree. For all the continuous variables we perform a simple shift-and-scale transformation.

The Matrix-Pareto is the only distribution which according to the Hill estimator will have the correct tail behaviour, and thus we consider it as a good target model for PH regression. However, we also consider a model which does not have the correct tail behaviour, namely the Matrix-Weibull distribution. The latter is convenient since it has a closed-form formula for mean prediction as per equation \eqref{weibull_mean_expression}. 

\begin{figure}[!htbp]
\centering
\includegraphics[clip, trim=3cm 3cm 2cm 2cm,width=.97\textwidth]{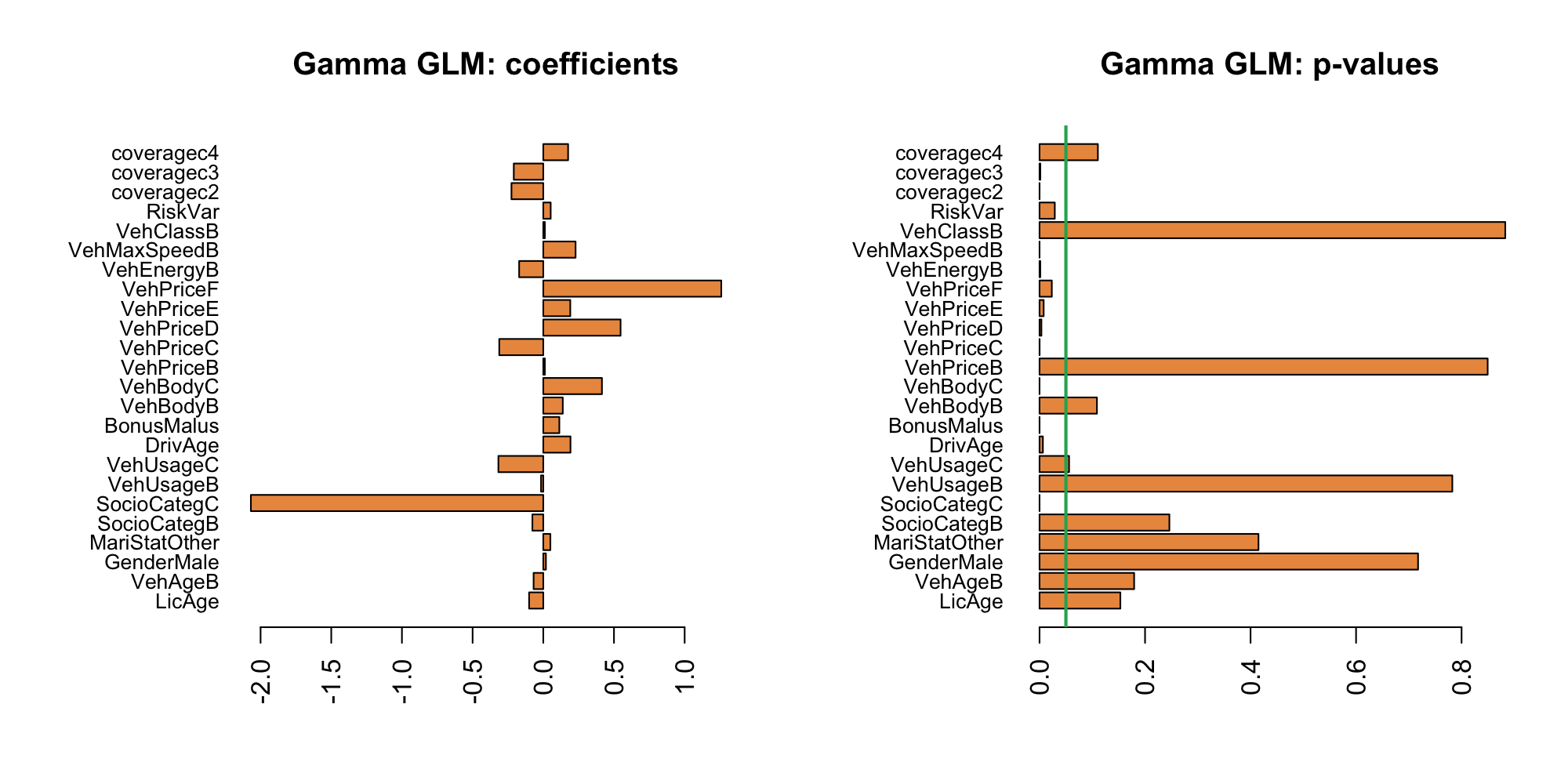}
\includegraphics[clip, trim=3cm 3cm 2cm 2cm,width=.97\textwidth]{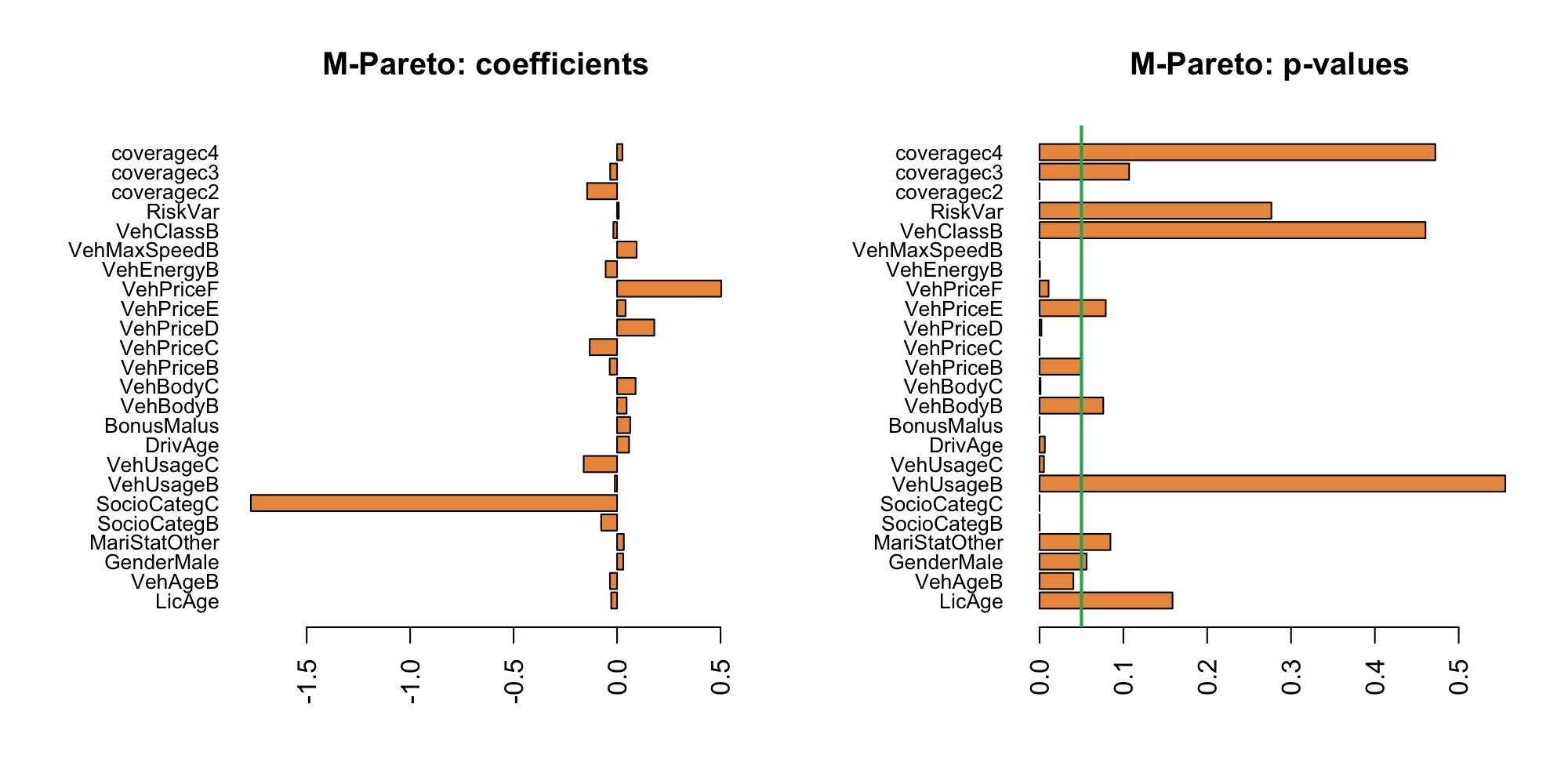}
\includegraphics[clip, trim=3cm 3cm 2cm 2cm,width=.97\textwidth]{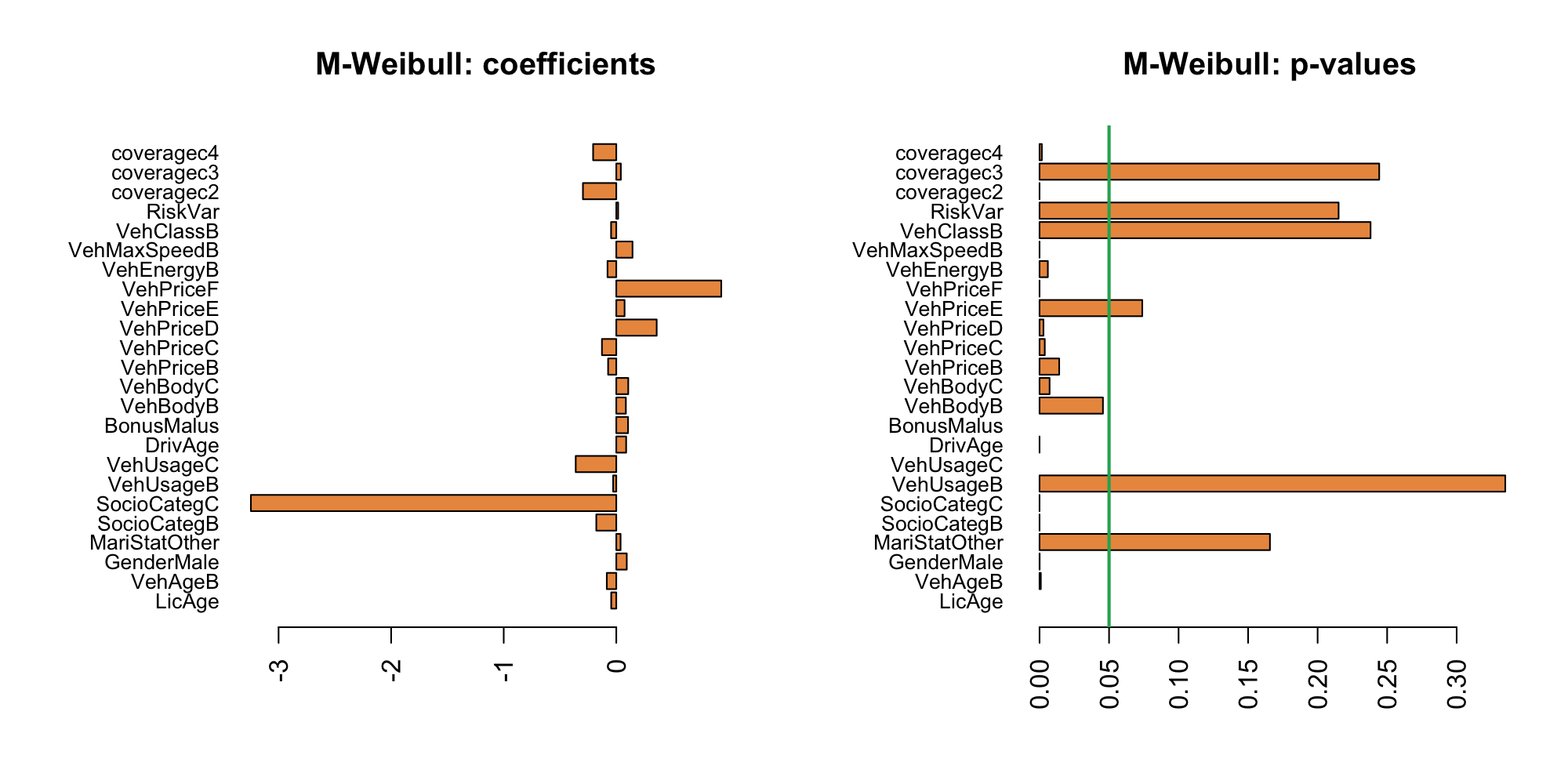}
\caption{Coefficients and p-values of IPH and GLM regression. For display: IPH coefficients multiplied by $-1$ and intercept of GLM  omitted.
}\label{fig:coefficients_mpl}
\end{figure}

The result of the fit using all the processed rating factors is given in Table \ref{table:coefficients_mpl} and Figure \ref{fig:coefficients_mpl}
, with all the coefficients of the $5$-dimensional IPH models multiplied by $-1$, to be comparable with the log-link Gamma GLM. The significance is obtained from a normal approximation using the implied Fisher matrix based on the Hessian matrix from the numerical optimization, {  as detailed in the goodness-of-fit subsection above}. When considering less covariates, { equation \ref{fisherinfo}} can be equivalently used, yielding very close results. The AIC and BIC are well below what the corresponding GLM can achieve, even if this is not a generous metric for IPH models. 

\begin{table}[!htbp]
\begin{center}
\caption{Summary for GLM and PH regression models for the freMPL dataset.}
\label{table:coefficients_mpl}
\begin{tabular}{l c c c}
\hline
& Gamma GLM & Pareto PH regression  & Weibull PH regression \\
\hline
Log Likelihood & $-60,368$   & $-59,464$   & $-59,446$   \\
Degrees of freedom & $26$   & $34$   & $34$   \\
AIC            & $120,788$   & $118,996$   & $\textbf{118,961}$   \\
BIC            & $120,966$   & $119,229$   & $\textbf{119,194}$   \\
Num. obs.      & $7,008$        & $7,008$        & $7,008$        \\
Loss-ratio  (pure)   & $101.03\%$      & $105.18\%$       & $101.13\%$      \\
\hline
\end{tabular}
\label{table:coefficients0}
\end{center}
\end{table}

We observe that the loss-ratio (where we understand as losses the claim severities, i.e. disregarding frequencies) is kept much better for the Weibull variant than for the Pareto, despite the latter being a better model for large claims. In practice, this inconvenience can be amended by regressing the losses with respect to the premium.  { Since estimation is done via MLE, it is not uncommon for a model with misspecified tail behaviour to globally perform better.} The fact that the Weibull PH regression performs almost as well as the GLM when predicting the mean is remarkable, since the former model does not specifically target averages, and the latter does. 

\begin{figure}[!htbp]
\centering
\includegraphics[clip, trim=0.5cm 1cm 2cm 1cm,width=1\textwidth]{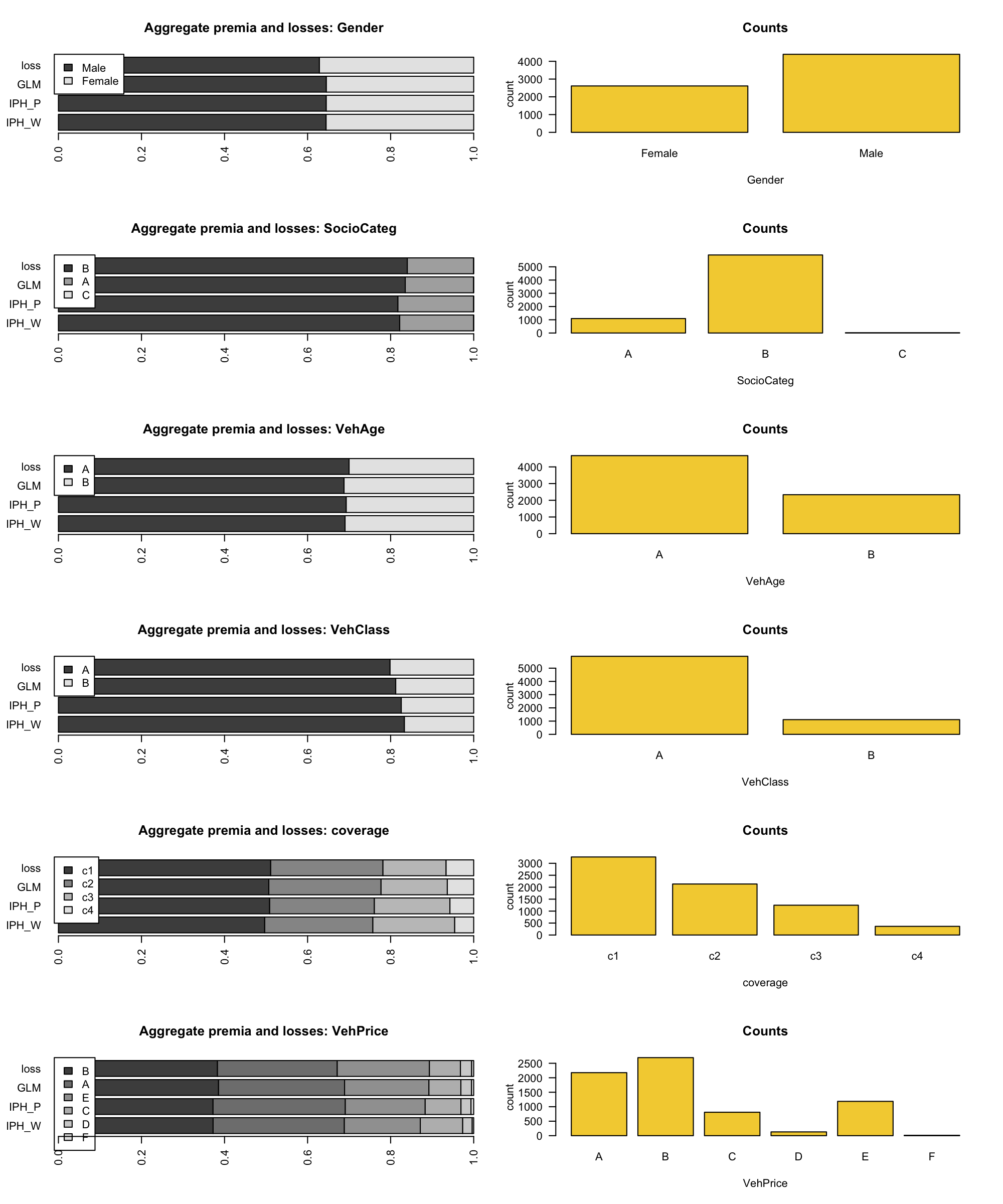}
\caption{Left panels: aggregate observed losses versus aggregate implied model premia (expected value), normalized to sum to one, for the Gamma GLM, {  and Pareto and }Weibull PH regressions; right panels: number of claims within each category (right).
}\label{fig:premia}
\end{figure}

{  In Figure \ref{fig:premia} we illustrate the behaviour of the aggregation of the implied mean predictions across all policies. We consider prediction from all three models, for selected rating factors and their categories, as computed by the general formula \eqref{mean_equation} in the Pareto case, and the explicit formula \eqref{weibull_mean_expression} in the Weibull case. Each of the aggregate predictions are normalized to sum to one for display purposes.} We again see that despite the Weibull PH regression not being {  specifically} designed for such task, it comes very close to the performance of the GLM.

\begin{figure}[!htbp]
\centering
\includegraphics[width=0.7\textwidth]{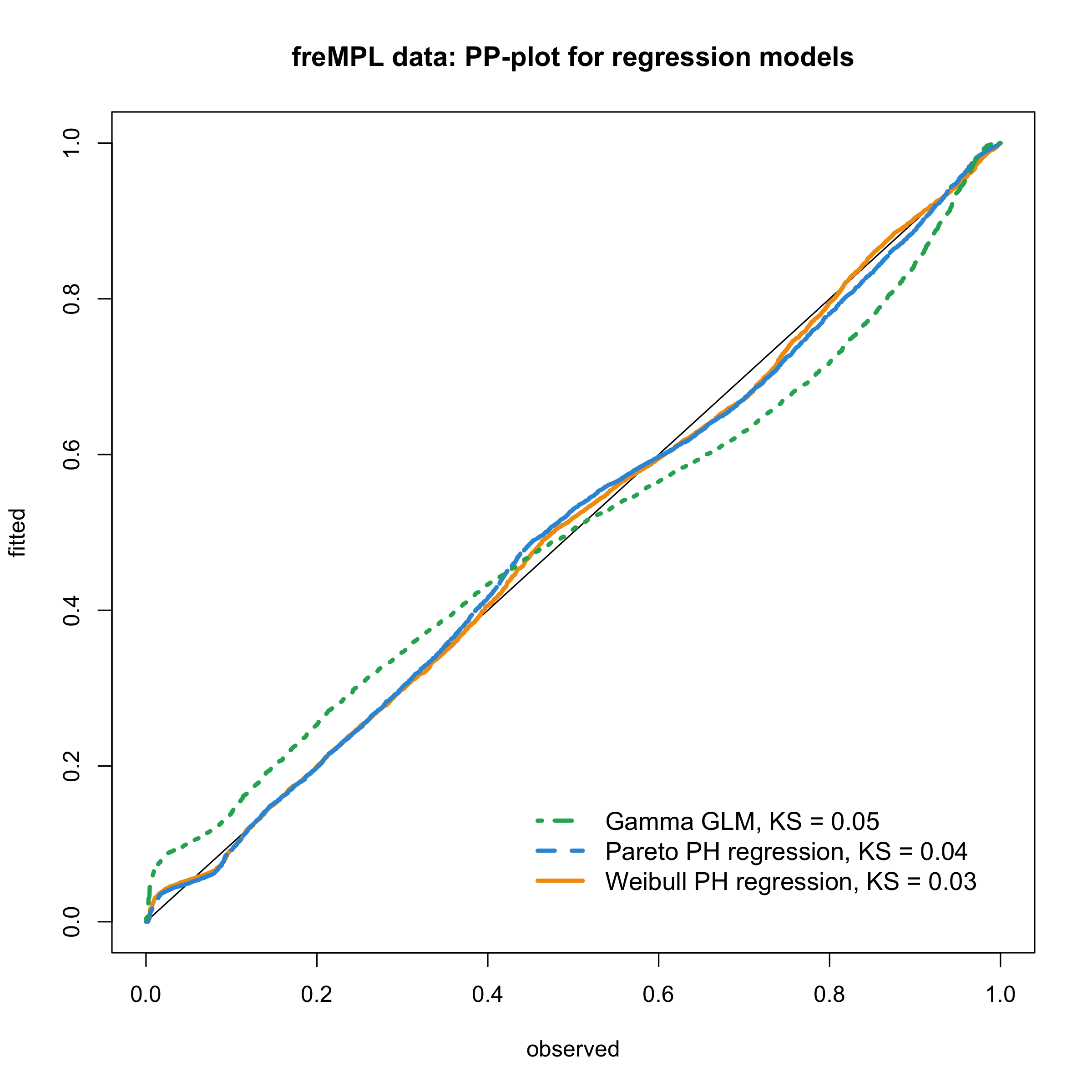}
\caption{ {  Ordered PIT's from equation \eqref{pp_specification} versus uniform order statistics for the French MPL dataset. { KS refers to the Kolmogorv-Smirnov statistic for testing uniformity}}
} \label{freMPL_reg}
\end{figure}

\begin{figure}[!htbp]
\centering
\includegraphics[clip, trim=0.2cm 1cm 1cm 1cm,width=1\textwidth]{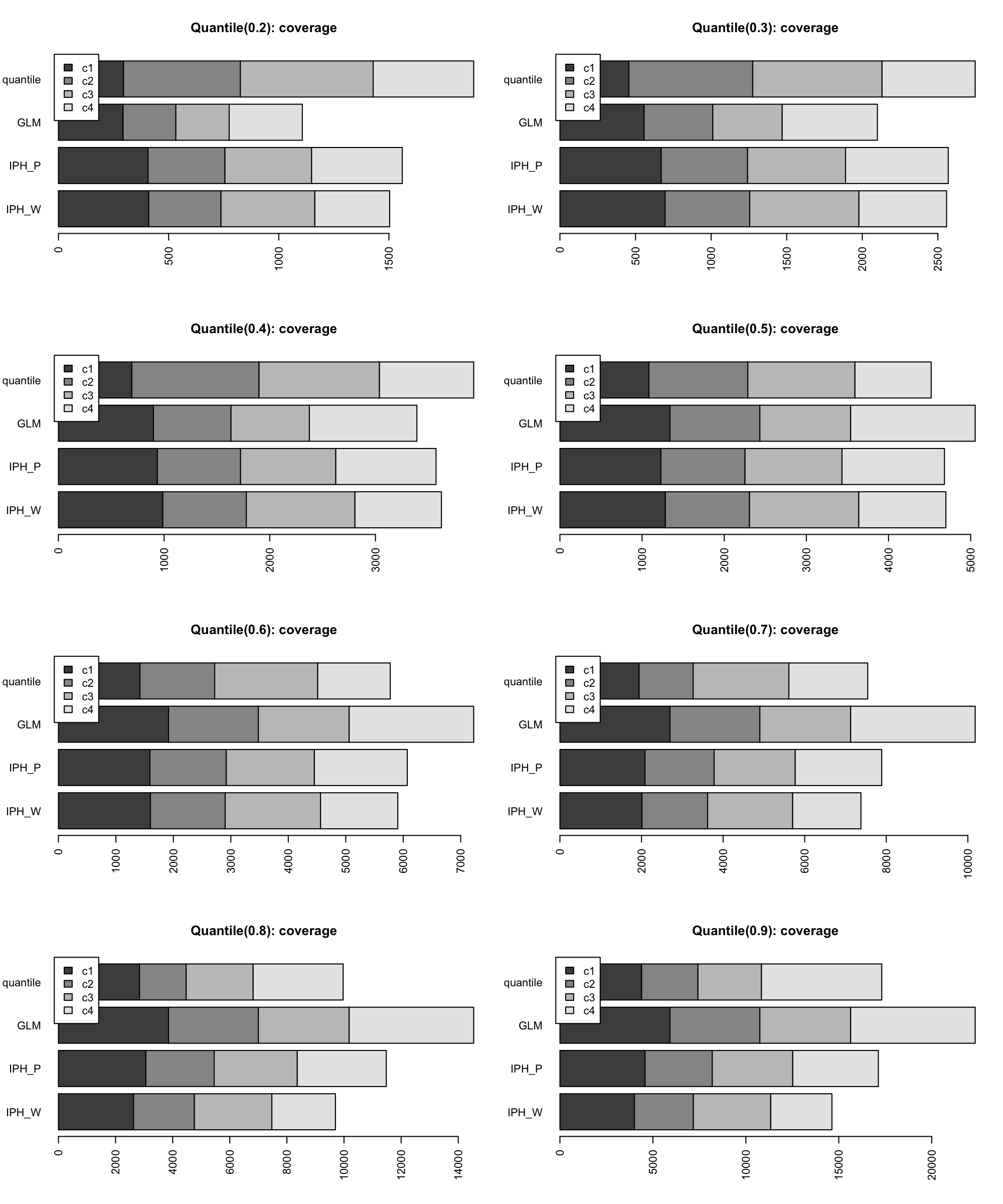}
\caption{Empirical quantiles by coverage category versus mean (accoss all other covariates) quantiles implied by the Gamma GLM, {  and Pareto and }Weibull PH regressions.
}\label{fig:quantiles}
\end{figure}

In Figure \ref{fig:quantiles} we illustrate the prediction of quantiles, as implied by the fitted conditional distribution functions of each model. We average the quantile estimates across policies corresponding to each of the rating factors and their categories. We observe that matrix distributions perform much better in this respect, and the correct specification of the Pareto tail can be appreciated in the bottom right panel, corresponding to the quantile at level $0.9$. Figure \ref{freMPL_reg} further supports the global estimation improvement in terms of the PP-plot goodness-of-fit diagnostic.

{  In terms of out-of-sample sample predictive performance, the mean square error (MSE) when holding out the first $20\%$ of the observations and training on the remaining $80\%$ is given by\footnote{Here, we have divided the MSE by $10^7$ for display purposes.} $3.66,\: 3.69,\: 3.70$, for the Gamma GLM, Pareto and Weibull PH regression models, respectively. In general, if the risk manager is solely interested in mean prediction, our experiments suggest that other data-driven methods may outperform PH regression models, although the advantage is usually small. However, the tables turn when considering the entire distribution of loss severities. Thus, a key takeaway here is that PH regression models should be used when the aim is to capture the entire distributional properties of claim severities correctly. In contrast, models specifically designed for mean prediction may be more appropriate if only the average is of interest.}

\section{Conclusion}\label{sec:conclusion}
We have presented a novel claim severities model based on PH distributions, which implicitly assumes an underlying  and unobservable  multi-state Markov structure. The inhomogeneity function is the key ingredient for translating the exponential tails of PH distributions to other tail domains, which is particularly useful for modeling of insurance data. We have shown that the use of a generalized EM algorithm performs effective estimation on the marginal distribution of severities, both with and without rating factors, the former being backed up by a simulation study. The flexibility of the PH regression model is particularly advantageous when data is multimodal, heavy- or light- tailed, and generally more heterogeneous than what the classical regression methods require. The practical implementation of PH, IPH and PH regression can be carried out using the \texttt{matrixdist} (\cite{matrixdist}) package in R.

Several questions remain open for further research, for instance the incorporation of multi-parameter PH regression models, or using IPH models to describe data exhibiting IBNR claims.  A faster implementation of the generalized EM algorithms is needed, mainly in the case when {  numerous categorical} rating factors are present, for widespread and systematic use of the PH regression model. Alternative estimation methods have not been explored, and could prove competitive to MLE. 

\textbf{Acknowledgement.} MB would like to acknowledge financial support from the Swiss National Science Foundation Project 200021\_191984.

\textbf{Declaration} MB declares no conflict of interest related to the current manuscript.

\bibliography{RPHV6.bib}

\newpage
\appendix
\section{Details on the EM algorithm for PH distributions}\label{apA}

Here we assume that $Z\sim\mbox{PH}( \bfp , \bfT ).$ Let $B_k$ be the number of times that the process $\{J_t\}_{t\geq0}$ starts in state $k$, $N_{ks}$ the total number of jumps from state $k$ to $s$, $N_k$ the number of times that we reach the absorbing state $p+1$ from state $k$ and let $Z_k$ be the total time that the underlying Markov jump process spends in state $k$ prior to absorption. These statistics are not recoverable from $Z$.
Given a sample of absorption times $\vect{z}$ the completely observed likelihood can be written in terms of these sufficient statistics as follows:
\begin{equation}
\mathcal{L}_c( \bfp , \bfT ;\vect{z})=
\prod_{k=1}^{p} {\pi_k}^{B_k} \prod_{k=1}^{p}\prod_{s\neq k} {t_{ks}}^{N_{ks}}e^{-t_{ks}Z_k}\prod_{k=1}^{p}{t_k}^{N_k}e^{-t_{k}Z_k},   
\end{equation}
which is seen to conveniently fall into the exponential family of distributions, and thus has explicit maximum likelihood estimators.

However, the full data is not observed, and hence we employ the expectation-maximization (EM) algorithm as an iterative way to obtain the maximum likelihood estimators. At each iteration the conditional expectations of the sufficient statistics $B_k$, $N_{ks}$, $N_k$ and $Z_k$ given the absorption times $\vect{z}$ are computed, commonly referred to as the E-step. Subsequently  $\mathcal{L}_c( \bfp , \bfT ,\vect{z})$ is maximised using the estimates of the sufficient statistics from the previous step, in this way obtaining $( \bfp , \bfT )$, commonly known as the M-step. The latter maximization is simple to perform because of the closed-form maximum likelihood estimators of exponential families. 

Below are the explicit formulas needed for the E- and M-steps for a sample of size $N$. We denote by ${\bfe}_k$ the column vector with all elements equal to zero besides the $k^{th}$ entry which is equal to one, i.e. the $k^{th}$ element of the canonical basis of $\mathbb{R}^d$.

\begin{enumerate} 
\item[ 1)]\textit{E-step, conditional expectations:} 
\begin{align*}
    \mathbb{E}(B_k\mid \mat{Z}=\vect{z})=\sum_{i=1}^{N} \frac{\pi_k {\bfe_k}^{ \mathsf{T}}\exp( \bfT z_i) \bft }{ \bfpi \exp( \bfT z_i) \bft }
\end{align*}
\begin{align*}
 \mathbb{E}(Z_k\mid \mat{Z}=\vect{z})=\sum_{i=1}^{N} \frac{\int_{0}^{x_i}{\bfe_k}^{ \mathsf{T}}\exp( \bfT (z_i-u)) \bft  \bfpi \exp( \bfT u)\vect{e}_kdu}{ \bfpi \exp( \bfT z_i) \bft }   
\end{align*}
\end{enumerate}

\begin{align*}
\mathbb{E}(N_{ks}\mid \mat{Z}=\vect{z})=\sum_{i=1}^{N}t_{ks} \frac{\int_{0}^{z_i}{\bfe_s}^{\mathsf{T}}\exp( \bfT (z_i-u)) \bft \, \bfpi \exp( \bfT u)\vect{e}_kdu}{ \bfpi \exp( \bfT z_i) \bft }
\end{align*}
\begin{align*}
\mathbb{E}(N_k\mid \mat{Z}=\vect{z})=\sum_{i=1}^{N} t_k\frac{ \bfpi  \exp( \bfT z_i){\bfe}_k}{ \bfpi \exp( \bfT z_i) \bft }.
\end{align*}
\begin{enumerate}

\item[2)] \textit{M-step, explicit maximum likelihood estimators:} 
\begin{align*}
\hat \pi_k=\frac{\mathbb{E}(B_k\mid \mat{Z}=\vect{z})}{N }, \quad \hat t_{ks}=\frac{\mathbb{E}(N_{ks}\mid \mat{Z}=\vect{z})}{\mathbb{E}(Z_{k}\mid \mat{Z}=\vect{z})}
\end{align*}
\begin{align*}
\hat t_{k}=\frac{\mathbb{E}(N_{k}\mid \mat{Z}=\vect{z})}{\mathbb{E}(Z_{k}\mid \mat{Z}=\vect{z})},\quad \hat t_{kk}=-\sum_{s\neq k} \hat t_{ks}-\hat t_k.
\end{align*}
We set $$ \hat{\bfp}=(\hat{\pi}_1, \dots,\: \hat{\pi}_p)^{\mathsf{T}},\quad \hat{\bfT} =\{{ \hat{t}}_{ks}\}_{k,s=1,2,\dots,p},\quad\hat{\bft}=( \hat{t}_1,\dots,\: \hat{t}_p)^{\mathsf{T}}.$$
\end{enumerate}

If we repeat the above two steps it can be shown that the likelihood increases at each iteration, and thus convergence to a possibly local maximum is guaranteed.

\end{document}